\documentclass[useAMS,usenatbib,usegraphicx]{mn2e}

\newcommand{\co}     {$^{13}$CO\,(1--0)}

\newcommand{\dvco}   {$^{12}$CO\,(1--0)}
\newcommand{\cs}     {CS\,(2--1)}

\newcommand{\kms}    {km\,s$^{-1}$}
\newcommand{\offsets}{($\Delta \alpha, \Delta \delta$)}
\newcommand{\vlsr}   {$V_{\rm lsr}$}

\newcommand{\hii}    {H\,{\sc{ii}}}
\newcommand{\trot}   {$T_{\rm rot}$}
\newcommand{\tkin}   {$T_{\rm kin}$}
\newcommand{\texit}  {$T_{\rm ex}$}

\newcommand{\tmb}    {$T_{\rm mb}$}
\newcommand{\nam}    {$N_{\rm NH_3}$}

\newcommand{\gnd}    {$n_{\rm H_2}$}
\newcommand{\am}     {NH$_3$}
\newcommand{\clone}  {S235\,East\,1}
\newcommand{\cltwo}  {S235\,East\,2}
\newcommand{\clcen}  {S235\,Central}
\newcommand{\cla}  {S235\,A}
\newcommand{\clb}  {S235\,B}

\newcommand{\rfrag}  {$R_{\rm frag}$}
\newcommand{\tfrag}  {$t_{\rm frag}$}
\newcommand{\mfrag}  {$M_{\rm frag}$}

\newcommand{\iss}  {$R_{\rm s }{(0)}$}

\title[Physical conditions in star forming regions around Sh2-235]{Physical conditions in star forming regions around S235}

\author[]{Kirsanova M. S.$^{1}$\thanks{E-mail:
kirsanova@inasan.ru}, Wiebe D. S.$^{1}$, Sobolev A. M.$^2$, Henkel C.$^{3,4}$, Tsivilev A. P.$^{5}$\\
$^{1}$Institute of Astronomy of the Russian Academy of Sciences, 48 Pyatnitskaya Str., Moscow 119017, Russia\\
$^{2}$Ural Federal University, 51 Lenin Str., Ekaterinburg 620051, Russia\\
$^{3}$Max Planck Instit\"{u}t f\"{u}r Radioastronomie, Auf dem H\"{u}gel 69, 53121 Bonn, Germany\\
$^{4}$Astron. Dept., King Abdulaziz University, P.O. Box 80203, Jeddah, Saudi Arabia \\
$^{5}$ Astro Space Center of Lebedev Physical Institute, Pushchino, Russian Academy of Sciences, Russia
}

\begin{document}

\date{Accepted \today, Received \today, in original form \today}

\pagerange{\pageref{firstpage}--\pageref{lastpage}} \pubyear{2011}

\maketitle

\label{firstpage}

\begin{abstract}
Gas density and temperature in star forming regions around Sh2-235 are derived from ammonia line observations. This information is used to evaluate formation scenarios and to determine evolutionary stages of the young embedded clusters \clone, \cltwo, and \clcen. We also estimate the gas mass in the embedded clusters and its ratio to the stellar mass. \clone\ appears to be less evolved than \cltwo\ and \clcen. In \clone\ the molecular gas mass exceeds that in the other clusters. Also, this cluster is more embedded in the parent gas cloud than the other two. Comparison with a theoretical model shows that the formation of these three clusters could have been stimulated by the expansion of the Sh2-235 \hii\ region (hereafter S235) via a collect-and-collapse process, provided the density in the surrounding gas exceeds $3\cdot10^3$~cm$^{-3}$, or via collapse of pre-existing clumps. The expansion of S235 cannot be responsible for star formation in the southern S235\,A-B region. However, formation of the massive stars in this region might have been triggered by a large-scale supernova shock. Thus, triggered star formation in the studied region may come in three varieties, namely collect-and-collapse and collapse of pre-existing clumps, both initiated by expansion of the local \hii\ regions, and triggering by an external large-scale shock. We argue that the \cla\ \hii\ region expands into a highly non-uniform medium with increasing density. It is too young to trigger star formation in its vicinity by a collect-and-collapse process.  There is an age spread inside the S235\,A-B region. Massive stars in the S235\,A-B region are considerably younger than lower mass stars in the same area. This follows from the estimates of their ages and the ages of associated \hii\ regions.

\end{abstract}


\section{Introduction}
\label{SEC:INTRO}

Among the various modes of star formation, triggered star formation attracts a special interest due to an increasing availability of multi-wavelength observations of expanding shells in the Milky Way \citep{bubble,bubblier}. The idea of self-stimulating star formation is quite old \citep{el77}. However, there are still only a few examples of star-forming regions, where triggering can be inferred more or less reliably \citep{example1,example2}. The reason is that expanding shells often have irregular shapes, so it is hard to deduce an evolutionary sequence based on morphology alone.

In our previous paper \citep[][hereafter Paper I]{s235_i} we presented arguments in favour of triggered star formation around an extended \hii\ region, Sh2-235 (hereafter S235), using both morphological and kinematical data. S235 is located in the Perseus Spiral Arm toward the anticentre of the Galaxy at $\alpha_{2000} = 05^{\rm h}41^{\rm m}33.\!\!^{\rm s}$8 and $\delta_{2000} = +35^{\circ}48'27.\!\!^{\prime\prime}0$. The region surrounding S235 is a site of active star formation. Young star clusters in its vicinity were recently investigated, e.g. by~\cite{allen_05}, \cite{klein_05}, \cite{kumar_06}, \cite{s235_i}, \cite{camargo_11}, and \cite{dewangan_11}. In Paper~I we argued that the formation of three clusters, located close to the edge of S235, could have been triggered by an expansion of the \hii\ region via a ``collect-and-collapse'' (C\&C) process due to gravitational instability in a dense material accumulated between the ionisation front and the shock wave. In the present paper we refer to these clusters as \clone, \cltwo, and \clcen, following the notation introduced in Paper~I. Their cross-identification can be found in~\citet{camargo_11}.

\begin{figure*}
\includegraphics[scale=0.9]{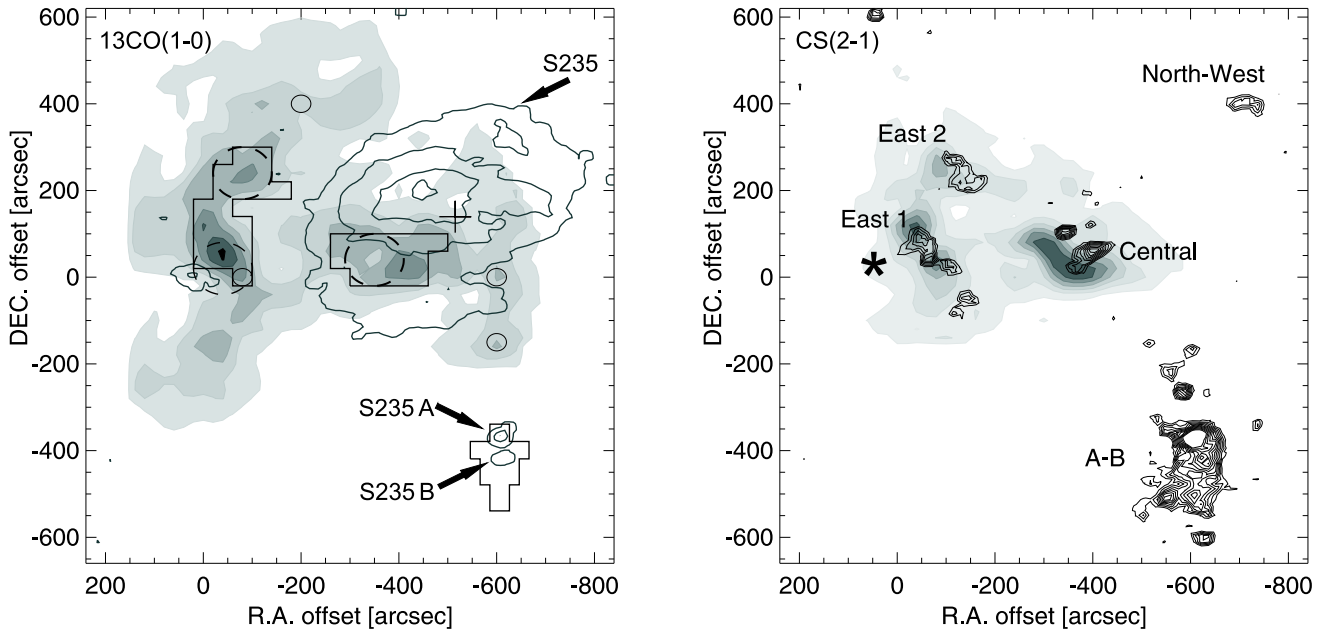}
\caption{Star forming regions around S235. Reference position is $\alpha_{2000} = 05^{\rm h}41^{\rm m}33.\!\!^{\rm s}$8 and $\delta_{2000} = +35^{\circ}48'27.\!\!^{\prime\prime}0$ in both panels. Distance to S235 is about 1.8\,kpc \citep{evans_81}, so that $1\arcmin$ corresponds to $\sim0.5$\,pc. Left panel: map of $^{13}$CO(1--0) emission from Paper~I. A cross marks location of the exciting star. Levels of greyscale correspond to 5, 15, 25, 35, 45 and 55~K~\kms. Contour levels show an optical image of the nebulae taken from the Digitized Sky Survey (DSS; POSS 2 red filter). The levels correspond to 30, 50, 70 and 90 per cent of the peak value. We removed point sources from the contour image to show the nebulae clearer. \hii\ regions S235 and S235\,A as well as the reflecting nebulae S235\,B are marked by arrows. Dashed circles indicate regions that were observed with the Pushchino 22-m telescope (RT-22). Thick black solid lines outline regions observed with the Effelsberg 100-m telescope. Sizes of dashed and small solid circles correspond to the full width at half power (FWHP) beam sizes of the RT-22 and the Effelsberg telescopes, respectively. Right panel: map of CS(2--1) emission (grey) and stellar density at 2~$\mu$m (contours) taken from Paper~I. The grey scale levels correspond to 1, 3, 5, 7, 9, and 11~K~\kms. Levels of stellar density are given from 20 to 60 stars~arcmin$^{-2}$ with a step of four stars~arcmin$^{-2}$. An asterisk marks location of IRAS source 05382+3547.}
\label{fig:general_view}
\end{figure*}

The inhomogeneously distributed molecular gas around the S235 \hii\ region \citep{heyer_96} complicates an analysis of the triggering scenario. In Paper~I arguments in favour of triggering were found on the basis of 1)~relative positions of the three embedded star clusters, dense molecular clumps, and the \hii\ region itself, and 2)~kinematics of molecular gas emitting \co\ and \cs\ lines. It was shown that the young star clusters \clone, \cltwo, and \clcen\ are embedded in dense molecular gas which forms a shell-like structure at the south-eastern side of S235. Each cluster is embedded in a separate dense clump, which is clearly visible on a \cs\ map. The morphology and location of the sources and the distribution of the molecular gas are shown in Fig.~\ref{fig:general_view}. Note that S235 is not completely surrounded by the dense shell. There is some gas emitting in \dvco\, and \co\, lines to the north and west from S235~\citep{heyer_96}. However, the \co\ brightness significantly decreases in these directions, so we had not observed that area during observing sessions described in Paper~I. Also, the column density of molecular gas located in front of the \hii\ region decreases to the south of S235 as this gas becomes transparent to the optical emission. In Paper~I we proposed that expansion of S235 triggered the formation of a dense shell and subsequent formation of three young embedded stellar clusters.

Three kinematic components were distinguished in Paper~I that may represent basic stages of interaction of young stellar objects with the surrounding medium. The first component that we refer to as primordial arguably comprises quiescent clumpy gas of the parent giant molecular cloud G174+2.5 that has not yet been disturbed by the expansion of S235. The second kinematic component represents gas that was entrained in motion and compressed by the expansion of the \hii\ region. This is the gas out of which the young star clusters have been formed. This gas still partially surrounds them. The third kinematic component supposedly corresponds to gas blown away by stellar winds from \cltwo\ and \clcen\ (see Fig.~\ref{fig:general_view}, right panel). It was proposed in Paper~I that these two clusters are, probably, more evolved than \clone.

The star forming region S235\,A-B encompasses both S235\,A and S235\,B nebulae and is located almost 10\arcmin\ to the south of \clcen. The S235\,A nebula is an expanding \hii\ region~\citep{isr_felli,felli_97}. \citet{boley_10} have shown that S235\,B is a reflection nebula. A water maser was found toward the region between \cla\ and \clb\ \citep{lo_75}  suggesting that massive star formation takes place there. Stellar mass estimates by \cite{dewangan_11} indicate that this water maser is associated with a young stellar object of high mass. The S235\,A-B region is not located in the compressed shell around the S235 \hii\ region. In Paper~I we suggested that star formation in S235\,A-B is not related to the expansion of S235.

This paper presents new observational data related to the star forming regions \clone, \cltwo, \clcen\ in the immediate vicinity of S235, and to regions around the \cla\ and \clb\ nebulae. The aim of this work is to determine physical conditions in these star forming regions. Ammonia emission was observed to estimate gas temperature and number density as this molecule is known to be both a thermometer and a densitometer for dense molecular gas~\citep{ho_83,walmsley_83}. The derived physical structure turns out to be consistent with our star formation scenario in this region and the inferred evolutionary stages of the young embedded star clusters.

\section{Observations and Data Reduction}
\label{SEC:OBSER}

Ammonia emission was mapped toward the \clone, \cltwo, and \clcen\ star clusters as well as toward S235\,A-B and primordial molecular gas at other locations.

\subsection{Pushchino (RT-22) Observations}
\label{subSEC:PSNOBS}
We first searched for ($J,K)=(1,1)$ and (2,2) emission ($\nu$ = 23.694496 and 23.722631\,GHz) toward three selected positions with the Pushchino 22-m telescope (RT-22, Astro Space Centre of Lebedev Physical Institute, Pushchino, Russian Academy of Sciences) in April 2006. Rest frequencies of observed ammonia transitions were taken from the catalogue of~\citet{lovas}. We used an ON-ON observing technique based on beam switching~\citep{berulis} with a beam throw of 10\arcmin\ at 23\,GHz. The main beam efficiency was 0.38 and the calibration uncertainty is estimated to be 10$\%$--15$\%$. We used the 50~MHz band of the Pushchino autocorrelator with 2048 channels. The spectral resolution was about 24~kHz. The full width at half power (FWHP) beam size of the 22-m telescope is about 2.6\arcmin\ which is larger than the size of the individual sources. Therefore these observations are suitable to estimate average parameters in the region. The ammonia lines were detected toward the three selected positions, shown in Fig.~\ref{fig:general_view} with dashed circles. The typical system temperature was about 150~K on an antenna temperature ($T_{\rm A}^*$) scale. Pointing was checked twice a day. The values of introduced corrections ranged from 0\arcsec\ to 20\arcsec.

\subsection{Effelsberg Observations}
\label{subSEC:EFF}

Emission from the ($J,K) = (1,1)$, (2,2), (3,3), and (4,4) ammonia transitions was mapped with the 100-m Effelsberg telescope in April and October 2007. We observed areas in star forming regions around S235 that are bright in the \cs\ emission and host young embedded star clusters. Long integrations were also performed toward locations of primordial gas. The regions observed in ammonia are shown in Fig.~\ref{fig:general_view} (left panel).

The FWHP of the Effelsberg telescope beam is about 40\arcsec\ at 23\,GHz. We used an 8192 channel correlator (AK90) split into 8 separate bands with a bandwidth of 10~MHz and 1024 channels. Thus we were able to obtain data on all four ammonia lines in two orthogonal linear polarizations simultaneously. The chosen backend configuration provided a channel spacing of 0.12~\kms\ (9.77~kHz). Observations were done in a position switching mode. The off-position was chosen to be located outside the giant molecular cloud G174+2.5~\citep{heyer_96} and has offset coordinates \offsets = (--15\arcmin,0\arcmin) with respect to our reference position of $\alpha_{2000}=05^{\rm h}41^{\rm m}33.\!\!^{\rm s}$8 and $\delta_{2000} = +35^{\circ}48'27.\!\!^{\prime\prime}0$. We checked the telescope pointing every hour. Corrections for azimuth and elevation were typically $<5\arcsec$, but sometimes $\sim10\arcsec$. The focus was checked after sunrise and sunset. Corrections for focusing were of the order of 1~mm. Data calibration was performed by observing the calibrators NGC7027, 3C123 and 3C286. A standard 22~GHz gain curve~\citep{stdgain} was used to account for a change of the telescope gain with elevation. Absolute and relative calibration uncertainties were estimated to be 15\% and 5\%, respectively. We spent about 12 minutes of on-time toward the majority of the positions. Positions of particular interest were measured twice. The map spacing was chosen to be 40\arcsec. The system temperature was 200--300\,K on a \tmb\ scale. The average noise level was 0.118~K on a \tmb\ scale.

Observations of the star forming region S235\,A-B were carried out in October~2007. Frontend and backend configurations were the same as for the previous observations, but the map spacing was reduced to 20\arcsec\, implying full sampling. The position of reference for the S235\,A-B spectra is
$\alpha_{2000} = 05^{\rm h}40^{\rm m}53.\!\!^{\rm s}$4 and $\delta_{2000} = +35^{\circ}41'48.^{\prime\prime}$0, the location of the HCO$^+$(1-0) emission peak~\citep{felli_s235ab}. It is situated to the south of the S235. The mapped area is also shown in Fig.~\ref{fig:general_view}.

Data obtained with the RT-22 and the Effelsberg telescope were reduced with the {\sc GILDAS} software\footnote{http://www.iram.fr/IRAMFR/GILDAS}. We used {\sc GILDAS} routine ``nh3(1,1)'' to estimate the optical depth $\tau_{1,1}$ of the ammonia $(J,K)=(1,1)$ line. Spectra of other ammonia transitions were fitted with the ``gauss'' routine to determine their intensity and width.

\section{Results}
\label{SEC:RESULT}

The standard interpretation of ammonia observations, described by \citet{stutzki}, was utilised to determine ammonia column density \nam\ and rotational temperature \trot. We calculated the gas number density (\gnd, essentially, number density of molecular hydrogen), using Eq.~2 from~\citet{ho_83}, toward positions with determined values of $\tau_{1,1}$ and NH$_3$ excitation temperature \texit. Rate coefficients for rotational excitation from \citet{danby} were used to determine the kinetic temperature \tkin\ at positions with known \trot. If data quality at a certain position was insufficient to determine $\tau_{1,1}$, i. e. the uncertainty of the value was higher than the value itself, we did not use corresponding $\tau_{1,1}$ and \texit\ estimates for further analysis.

\begin{figure}
\includegraphics[scale=0.45]{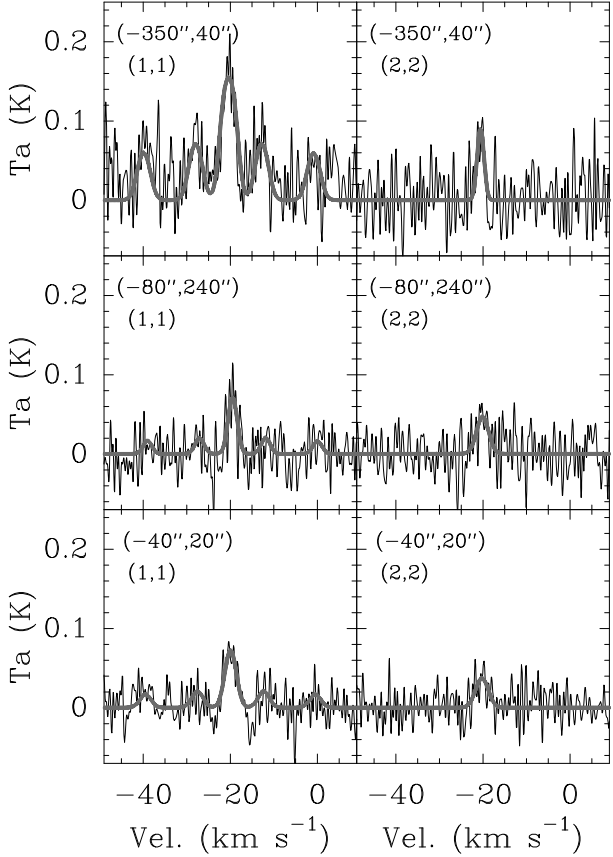}
\caption{Spectra of \am(1,1) and \am(2,2) emission toward selected positions, obtained with the RT-22 telescope. The two lines were observed simultaneously within a single band. Solid lines show spectra fitted with the GILDAS ``nh3(1,1)'' and ``gauss'' routines for \am(1,1) and \am(2,2) lines, respectively. All spectra are Hanning-smoothed. For the reference position, see Fig~\ref{fig:general_view}.}
\label{fig:rt22_spectra}
\end{figure}

Ammonia spectra, measured with the Pushchino 22-m telescope, are shown in Fig.~\ref{fig:rt22_spectra}. Satellites of the \am(1,1) line are seen toward all observed positions, while the hyperfine structure of the \am(2,2) line is not pronounced. The quality of our spectra allows estimating \tkin\ toward all young clusters under study. The values are presented in Table~\ref{tab:rt22_res}. The gas number density could be determined toward \clcen\, because we could determine \texit\ toward this position. Numbers in parentheses in all tables show the values of uncertainties. Table~\ref{tab:rt22_res} indicates that \tkin\ in the direction of \clone\ is considerably lower than \tkin\ toward \cltwo\ and \clcen. As it was mentioned already, results of observations with RT-22 represent physical parameters averaged over a large area.

\begin{table*}
\caption{Observational results and physical parameters derived from RT-22 telescope data. Reference position is $\alpha_{2000} = 05^{\rm h}41^{\rm m}33.\!\!^{\rm s}$8 and $\delta_{2000} = +35^{\circ}48'27.\!\!^{\prime\prime}0$.}
\begin{tabular}{cccccccc}
\hline
Offset     & (J,K)  & \tmb        & Width       & \vlsr        & $\tau$ & \tkin  & $n$\\
($\arcsec$,$\arcsec$)    &        & (K)         & (\kms)      & (\kms)       &        & (K)    & (10$^3$ cm$^{-3}$)\\
\hline
 {\bf \clone} \\
(--40, 20) & (1,1)  & 0.08 (0.02) & 3.04 (0.29) &--20.00 (0.16) & --     &  23.38 (8.44) & --\\
           & (2,2)  & 0.04 (0.02) & 3.47 (0.72) &--20.31 (0.38) & --     &       \\
 {\bf \cltwo} \\
(--80,240)   & (1,1)  & 0.08 (0.03) & 2.43 (0.47) &--19.53 (0.20) & --     & 32.25 (20.16)  & --\\
           & (2,2)  & 0.05 (0.03) & 3.29 (0.98) &--20.15 (0.38) & --     &     \\
 {\bf \clcen} \\
(--350,40) & (1,1)  & 0.15 (0.03) & 2.24 (0.26) &--20.40 (0.10) & 0.98 (0.64) & 30.24 (11.58) & 0.49\\
           & (2,2)  & 0.09 (0.03) & 1.88 (0.42) &--20.62 (0.21) & --     &       \\
\hline
\end{tabular}
\label{tab:rt22_res}
\end{table*}

\begin{figure}
\includegraphics[scale=0.45]{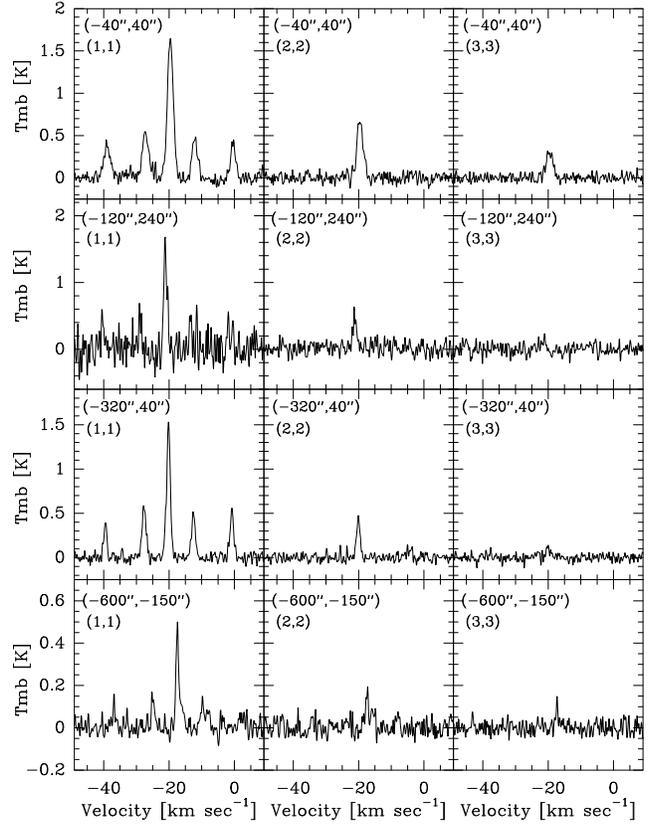}
\caption{Spectra of \am(1,1) (left), \am(2,2) (middle) and \am(3,3) (right) emission toward three star-forming clumps (three rows of spectra from the top) and primordial gas (bottom) of the S235 molecular complex. The spectra toward the offsets (--40\arcsec,40\arcsec), (--120\arcsec,240\arcsec) and (--320\arcsec,40\arcsec) correspond to the \clone, \cltwo\ and \clcen\ clumps, respectively. The spectra are Hanning-smoothed.}
\label{ex_spectra}
\end{figure}

With the Effelsberg telescope, we detected \am(1,1) and \am(2,2) lines toward almost all selected positions, while the (3,3) transition was detected in 10 of 43 positions. The (4,4) line remained undetected. Representative \am(1,1), \am(2,2), and \am(3,3) lines are shown in Fig.~\ref{ex_spectra}. The hyperfine structure of \am(1,1) emission provides a possibility to determine the optical depth of the line. The \am(1,1) emission lines are optically thin in most of the observed positions.  Observational results obtained with the Effelsberg telescope are presented in Table~\ref{tab:obs_res}. The signal-to-noise ratio allowed estimating \tkin\ toward 37 positions. The gas number density was estimated in 15 positions. Several positions were identified where intensity ratios of hyperfine components differ from their LTE values. This information can be used to put some additional constraints on the cloud structure (e.g., \S3.2).

\begin{table*}
\caption{Parameters of ($J,K$)=(1,1), (2,2) and (3,3) ammonia lines measured with the Effelsberg 100-m telescope.}
\begin{tabular}{cccccc}
\hline
Offset             & ($J,K$) & \tmb        & Width       & \vlsr         & $\tau$ \\
($\arcsec$,$\arcsec$)            &       & (K)         & (\kms)      & (\kms)        &        \\
\hline
\multicolumn{6}{l} {{\bf \clone}, ref. pos. $\alpha_{2000} = 05^{\rm h}41^{\rm m}33.\!\!^{\rm s}$8, $\delta_{2000} = +35^{\circ}48'27.\!\!^{\prime\prime}0$} \\
0, 160             & (1,1) & 0.55 (0.17) & 2.14 (0.28) &--21.39 (0.10) & --     \\
                   & (2,2) & $<$  0.20   & --          & --            & --     \\
                   & (3,3) & $<$  0.20   & --          & --            & --     \\
0, 120             & (1,1) & 0.77 (0.18) & 2.34 (0.25) &--20.90 (0.11) & --     \\
                   & (2,2) & $< 0.18$    & --          & --            & --     \\
                   & (3,3) & $< 0.20$    & --          & --            & --     \\
0, 80              & (1,1) & 0.77 (0.20) & 1.95 (0.21) &--19.95 (0.09) & --     \\
                   & (2,2) & 0.17 (0.16) & 3.78 (1.11) &--18.97 (0.63) & --     \\
                   & (3,3) & $< 0.20$    & --          & --            & --     \\
0, 40              & (1,1) & 0.50 (0.19) & 1.52 (0.22) &--19.37 (0.11) & --     \\
                   & (2,2) & $< 0.18$    & --          & --            & --     \\
                   & (3,3) & $< 0.20$    & --          & --            & --     \\
--40, 160          & (1,1) & 0.52 (0.18) & 1.06 (0.43) &--20.70 (0.09) & --     \\
                   & (2,2) & $< 0.19$    & --          & --            & --     \\
                   & (3,3) & $< 0.20$    & --          & --            & --     \\
--40, 120          & (1,1) & 1.16 (0.17) & 2.30 (0.14) &--20.87 (0.07) & 0.16 (0.09) \\
                   & (2,2) & 0.41 (0.08) & 2.42 (0.21) &--20.78 (0.09) & --     \\
                   & (3,3) & 0.21 (0.08) & 3.59 (0.38) &--20.17 (0.17) & --     \\
--40, 80           & (1,1) & 2.38 (0.16) & 1.95 (0.08) &--20.20 (0.03) & 0.41 (0.19) \\
                   & (2,2) & 0.37 (0.08) & 2.22 (0.68) &--20.10 (0.20) & --     \\
                   & (3,3) & $<$  0.24   &--           &--             &--      \\
--40, 40           & (1,1) & 1.87 (0.06) & 1.97 (0.03) &--19.62 (0.01) & 0.29 (0.08) \\
                   & (2,2) & 0.67 (0.06) & 2.21 (0.10) &--19.56 (0.04) & 0.10 (0.07) \\
	               & (3,3) & 0.31 (0.09) & 2.43 (0.30) &--19.57 (0.13) & --          \\
--80, 120          & (1,1) & 0.23 (0.21) & 2.69 (1.08) &--14.31 (0.51) & --      \\
                   & (2,2) & $< 0.33$    & --          & --            & --     \\
                   & (3,3) & $< 0.20$    & --          & --            & --     \\
--80, 80           & (1,1) & 0.53 (0.18) & 1.54 (0.31) &--19.78 (0.11) & --     \\
                   & (2,2) & $< 0.21$    & --          & --            & --     \\
                   & (3,3) & $< 0.20$    & --          & --            & --     \\
--80, 40           & (1,1) & 1.10 (0.11) & 1.73 (0.07) & -19.21 (0.03) & 0.29 (0.21) \\
                   & (2,2) & 0.45 (0.08) & 1.81 (0.19) & -19.14 (0.07) & --     \\
                   & (3,3) & $<$  0.24   &--           &--             & --     \\
--80, 0            & (1,1) & 1.27 (0.18) & 1.27 (0.09) & -19.06 (0.04) & --     \\
                   & (2,2) & 0.49 (0.08) & 1.11 (0.23) & -19.20 (0.10) & --     \\
                   & (3,3) & $< 0.24$    &--           &--             &--      \\
\multicolumn{6}{l} {{\bf \cltwo}, ref. pos. $\alpha_{2000} = 05^{\rm h}41^{\rm m}33.\!\!^{\rm s}$8, $\delta_{2000} = +35^{\circ}48'27.\!\!^{\prime\prime}0$ }\\
--40, 240          & (1,1) & 0.67 (0.19) & 1.68 (0.19) &--21.14 (0.09) & --     \\
                   & (2,2) & $< 0.17$    & --          & --            & --     \\
                   & (3,3) & $< 0.20$    & --          & --            & --     \\
--40, 200          & (1,1) & 0.61 (0.06) & 1.91 (0.10) & -21.24 (0.04) & --     \\
                   & (2,2) & 0.19 (0.05) & 1.58 (0.36) & -21.35 (0.12) & --     \\
	               & (3,3) & 0.07 (0.03) & 2.22 (0.49) & -20.84 (0.16) & --     \\
--80, 200          & (1,1) & 0.55 (0.11) & 1.78 (0.16) & -20.96 (0.06) & 0.60 (0.44) \\
                   & (2,2) & 0.23 (0.08) & 1.89 (0.37) & -20.66 (0.14) & --     \\
                   & (3,3) & $< 0.10$    &--           &--             &--      \\
--80, 240          & (1,1) & 1.33 (0.34) & 1.27 (0.12) & -21.28 (0.12) & --     \\
                   & (2,2) & 0.37 (0.14) & 1.31 (0.31) & -21.32 (0.10) & --     \\
                   & (3,3) & $< 0.30$    &--           &--             &--      \\
--80, 280          & (1,1) & 0.97 (0.10) & 1.48 (0.09) & -21.37 (0.03) & --     \\
                   & (2,2) & 0.30 (0.06) & 1.18 (0.21) & -21.10 (0.08) & --     \\
                   & (3,3) & $< 0.18$    &--           &--             &--      \\
--120, 200         & (1,1) & 0.90 (0.16) & 1.34 (0.18) & -21.13 (0.07) & --     \\
                   & (2,2) & 0.48 (0.07) & 0.84 (0.18) & -20.96 (0.05) & --     \\
                   & (3,3) & $< 0.15$    &--           &--             &--      \\
--120, 240         & (1,1) & 1.58 (0.30) & 1.19 (0.12) & -21.10 (0.04) & --     \\
                   & (2,2) & 0.49 (0.06) & 1.44 (0.20) & -21.19 (0.08) & --     \\
                   & (3,3) & 0.14 (0.07) & 1.61 (0.98) & -21.39 (0.38) & --     \\
--120, 280         & (1,1) & 1.13 (0.16) & 1.47 (0.15) & -21.00 (0.05) & --     \\
                   & (2,2) & $< 0.16 $   &--           &--             &--      \\
                   & (3,3) & $< 0.16 $   &--           &--             &--      \\
--160, 200         & (1,1) & 0.18 (0.07) & 1.42 (0.40) & -20.99 (0.16) & --     \\
                   & (2,2) & 0.17 (0.06) & 1.56 (0.25) & -20.90 (0.10) & --     \\
	               & (3,3) & 0.16 (0.04) & 0.44 (0.25) & -20.90 (0.06) & --     \\
\hline
\end{tabular}
\label{tab:obs_res}
\end{table*}

\begin{table*}
\contcaption{Parameters of detected ($J,K$)=(1,1), (2,2) and (3,3) ammonia lines.}
\begin{tabular}{cccccc}
\hline
Offset             & ($J,K$) & \tmb        & Width       & \vlsr         & $\tau$ \\
($\arcsec$,$\arcsec$)            &       & (K)         & (\kms)      & (\kms)        &        \\
\hline
\multicolumn{6}{l} {{\bf \clcen}, ref. pos. $\alpha_{2000} = 05^{\rm h}41^{\rm m}33.\!\!^{\rm s}$8, $\delta_{2000} = +35^{\circ}48'27.\!\!^{\prime\prime}0$}\\
     --280, 80     & (1,1) & 0.75 (0.13) & 1.44 (0.12) & -19.97 (0.05) & --     \\
                   & (2,2) & 0.33 (0.09) & 1.47 (0.34) & -19.90 (0.11) & --     \\
	               & (3,3) & 0.18 (0.11) & 0.98 (0.42) & -19.92 (0.20) & --     \\
--280, 40          & (1,1) & 0.35 (0.10) & 2.84 (0.12) & -20.26 (0.11) & 0.98 (0.67) \\
                   & (2,2) & 0.23 (0.11) & 1.24 (0.43) & -20.81 (0.19) & --     \\
                   & (3,3) & $<$  0.20   &--           &--             &--      \\
--320, 80          & (1,1) & 1.12 (0.17) & 1.46 (0.10) & -20.08 (0.04) & 1.05 (0.35) \\
                   & (2,2) & 0.50 (0.05) & 1.54 (0.17) & -20.01 (0.07) &--\\
                   & (3,3) &  $<$  0.15  &--           &--             &--\\
--320, 40          & (1,1) & 1.45 (0.08) & 1.13 (0.03) &--20.18 (0.01) & 0.90 (0.12) \\
                   & (2,2) & 0.45 (0.07) & 1.24 (0.09) &--20.08 (0.04) &--\\
	               & (3,3) & 0.11 (0.06) & 1.56 (0.48) &--20.00 (0.19) &--\\
--320, 0           & (1,1) & 0.35 (0.17) & 2.03 (0.50) &--18.95 (0.24) &--\\
                   &       & 0.38 (0.17) & 1.40 (0.35) &--21.01 (0.17) &--\\
                   & (2,2) & 0.16 (0.15) & 1.41 (0.66) &--20.27 (0.35) &--\\
                   & (3,3) & $<$  0.16   &--           &--             &--\\
--360, 80          & (1,1) & 0.53 (0.17) & 2.41 (0.28) &--20.23 (0.12) &--\\
                   & (2,2) & $<0.18 $    &--           &--             &--\\
                   & (3,3) & $<0.18 $    &--           &--             &--\\
--360, 40          & (1,1) & 0.53 (0.11) & 2.24 (0.19) & -20.42 (0.08) &--\\
                   & (2,2) & 0.25 (0.09) & 2.22 (0.28) & -20.26 (0.11) &--\\
                   & (3,3) & $<$  0.10   &--           &--             &--\\
--360, 0           & (1,1) & 0.56 (0.16) & 0.82 (0.19) &--18.65 (0.07) &--\\
                   &       & 0.28 (0.16) & 2.41 (0.58) &--20.78 (0.25) &--\\
                   & (2,2) & $<0.14 $    &--           &--             &--\\
                   & (3,3) & $<0.17 $    &--           &--             &--\\
--400, 80          & (1,1) & 0.42 (0.17) & 3.06 (0.38) &--21.52 (0.17) &--\\
                   & (2,2) & $<0.17 $    &--           &--             &--\\
                   & (3,3) & $<0.20 $    &--           &--             &--\\
--400, 40          & (1,1) & 0.93 (0.13) & 1.66 (0.17) &--21.56 (0.06) &--\\
                   & (2,2) & $<0.18 $    &--           &--             &--\\
                   & (3,3) & $<0.20 $    &--           &--             &--\\
--400, 0           & (1,1) & 0.79 (0.20) & 2.55 (0.23) &--20.80 (0.09) &0.61 (0.45)\\
                   & (2,2) & 0.20 (0.18) & 1.44 (0.51) &--21.17 (0.31) &--\\
                   & (3,3) & $<0.20 $    &--           &--             &--\\
--440, 80          & (1,1) & 0.37 (0.16) & 1.54 (0.42) &--21.73 (0.15) &--\\
                   & (2,2) & $<0.11$     &--           &--             &--\\
                   & (3,3) & $<0.11$     &--           &--             &--\\
--440, 40          & (1,1) & 0.61 (0.16) & 1.87 (0.21) &--21.70 (0.09) &--\\
                   & (2,2) & $<0.09$     &--           &--             &--\\
                   & (3,3) & $<0.15$     &--           &--             &--\\
--440, 0           & (1,1) & 0.54 (0.16) & 1.46 (0.20) &--21.00 (0.07) &1.07 (0.69)\\
                   & (2,2) & 0.09 (0.06) & 1.68 (0.80) &--20.70 (0.33) &--\\
                   & (3,3) & $<0.18$     &--           &--             &--\\
--480, 80          & (1,1) & 0.98 (0.08) & 1.07 (0.06) & -21.84 (0.02) & 0.52 (0.17) \\
                   & (2,2) & 0.50 (0.08) & 1.34 (0.12) & -21.68 (0.05) &--\\
	               & (3,3) & 0.12 (0.10) & 2.05 (1.16) & -21.72 (0.42) &--\\
\multicolumn{6}{l} {{\bf Primordial gas}, ref. pos. $\alpha_{2000} = 05^{\rm h}41^{\rm m}33.\!\!^{\rm s}$8, $\delta_{2000} = +35^{\circ}48'27.\!\!^{\prime\prime}0$}\\
--600, --150       & (1,1) & 0.46 (0.05) & 1.10 (0.09) & -17.43 (0.03) &--\\
                   & (2,2) & 0.11 (0.05) & 0.64 (0.16) & -17.29 (0.08) &--\\
	               & (3,3) & 0.15 (0.03) & 0.51 (0.22) & -17.29 (0.08) &--\\
--600, 0           & (1,1) & 0.32 (0.09) & 1.69 (0.20) & -16.82 (0.09) &--\\
                   & (2,2) & 0.17 (0.09) & 1.36 (0.34) & -17.11 (0.14) &--\\
	               & (3,3) & $<$ 0.10    &  --         &      --       &--\\
     --200, 400    & (1,1) & 0.21 (0.08) & 0.67 (0.27) &--20.72 (0.08) &--\\
                   & (2,2) & $< 0.06$    & --          &--             &--\\
                   & (3,3) & $< 0.08$    & --          &--             &--\\
\hline
\end{tabular}
\end{table*}

\begin{table*}
\contcaption{Parameters of detected ($J,K$)=(1,1), (2,2) and (3,3) ammonia lines.}
\begin{tabular}{cccccc}
\hline
Offset             & ($J,K$) & \tmb        & Width       & \vlsr         & $\tau$ \\
($\arcsec$,$\arcsec$)            &       & (K)         & (\kms)      & (\kms)        &        \\
\hline
   &       &             &             &               &\\
\multicolumn{6}{l} {{\bf {\bf S235\,A-B}}, ref. pos. $\alpha_{2000} = 05^{\rm h}40^{\rm m}53.\!\!^{\rm s}$4, $\delta_{2000} = +35^{\circ}41'48.\!\!^{\prime\prime}0$}\\
0,40               & (1,1) & 0.47 (0.22) & 2.25 (0.81) &--16.42 (0.28) &-- \\
                   & (2,2) & $< 0.18 $&-- &-- &-- \\
                   & (3,3) & $< 0.18 $&-- &-- &-- \\
0,20               & (1,1) & 1.12 (0.22) & 2.12 (0.21) &--16.61 (0.10) & 0.67 (0.54)\\
                   & (2,2) & 0.56 (0.17) & 1.95 (0.37) &--16.29 (0.15) &--\\
                   & (3,3) & 0.79 (0.40) & 1.84 (0.74) &--16.60 (0.27) &--\\
--40,0             & (1,1) & 0.05 (0.03) & 2.36 (0.81) &--23.38 (0.35) &--\\
                   &       & 0.08 (0.03) & 2.42 (0.55) &--16.58 (0.23) &--\\
                   & (2,2) & $< 0.13 $&-- &-- &-- \\
                   & (3,3) & $< 0.30 $&-- &-- &-- \\
0,0                & (1,1) & 2.07 (0.13) & 2.17 (0.07) &--16.83 (0.03) &0.33 (0.17)\\
                   & (2,2) & 1.50 (0.21) & 2.35 (0.15) &--16.90 (0.07) &--\\
                   & (3,3) & 0.73 (0.09) & 2.70 (0.27) &--16.83 (0.11) &--\\
40,0               & (1,1) & 0.70 (0.21) & 1.96 (0.38) &--15.94 (0.15) &--\\
                   & (2,2) & 0.23 (0.15) & 2.85 (0.80) &--16.28 (0.38) &--\\
                   & (3,3) & $< 0.30 $&-- &-- &-- \\
20,--20            & (1,1) & 2.71 (0.24) & 2.14 (0.10) &--16.70 (0.05) & 0.276 ( 0.223) \\
                   & (2,2) & 0.84 (0.20) & 2.01 (0.26) &--16.72 (0.12) &--\\
                   & (3,3) & 0.72 (0.37) & 3.06 (0.70) &--16.29 (0.31) &--\\
0,--20             & (1,1) & 3.25 (0.20) & 1.98 (0.08) &--16.84 (0.03) &--\\
                   & (2,2) & 0.90 (0.13) & 2.26 (0.29) &--16.87 (0.10) &--\\
                   & (3,3) & 0.78 (0.25) & 2.35 (0.51) &--17.16 (0.24) &--\\
--20,--20          & (1,1) & 0.55 (0.21) & 2.29 (0.60) &--16.05 (0.31) &--\\
                   & (2,2) & 0.40 (0.20) & 4.29 (1.04) &--15.99 (0.40) &--\\
                   & (3,3) & 0.39 (0.38) & 2.10 (1.21) &--17.17 (0.62) &--\\
0,--40             & (1,1) & 3.35 (0.24) & 1.83 (0.08) &--16.56 (0.03) &--\\
                   & (2,2) & 0.66 (0.17) & 2.96 (0.43) &--16.03 (0.17) &--\\
                   & (3,3) & 0.44 (0.32) & 5.88 (1.25) &--15.15 (0.63) &--\\
0,--60             & (1,1) & 2.84 (0.23) & 2.29 (0.10) &--16.07 (0.04) &--\\
                   & (2,2) & 0.61 (0.18) & 1.68 (0.32) &--16.10 (0.14) &--\\
                   & (3,3) & 0.38 (0.26) & 2.02 (0.79) &--16.04 (0.47) &--\\
--20,--60          & (1,1) & 1.70 (0.20) & 2.20 (0.17) &--15.91 (0.08) &--\\
                   & (2,2) & 0.77 (0.30) & 1.84 (0.38) &--16.36 (0.18) &--\\
                   & (3,3) & 0.50 (0.27) & 2.62 (0.99) &--15.18 (0.45) &--\\
20,--60            & (1,1) & 2.40 (0.25) & 1.96 (0.13) &--16.24 (0.05) &--\\
                   & (2,2) & 0.72 (0.33) & 2.64 (0.89) &--16.33 (0.28) &--\\
                   & (3,3) & 0.73 (0.33) & 2.28 (0.69) &--16.44 (0.24) &--\\
0,--80             & (1,1) & 2.61 (0.19) & 2.16 (0.08) &--16.01 (0.03) &0.48 (0.18) \\
                   & (2,2) & 0.48 (0.13) & 3.03 (0.34) &--15.63 (0.16) &--\\
                   & (3,3) & 0.45 (0.24) & 2.48 (0.86) &--15.94 (0.30) &--\\
0,--100            & (1,1) & 1.74 (0.21) & 2.49 (0.17) &--15.95 (0.07) &--\\
                   & (2,2) & 0.65 (0.27) & 3.80 (0.68) &--15.58 (0.28) &--\\
                   & (3,3) & $< 0.20 $&-- &-- &-- \\
0,--120            & (1,1) & 0.47 (0.13) & 3.74 (0.58) &--16.01 (0.27) &--\\
                   & (2,2) & 0.20 (0.17) & 2.00 (0.80) &--17.24 (0.39) &--\\
                   & (3,3) & $< 0.18 $&-- &-- &-- \\
\hline
\end{tabular}
\end{table*}

\begin{figure}
\includegraphics[scale=0.5,angle=270]{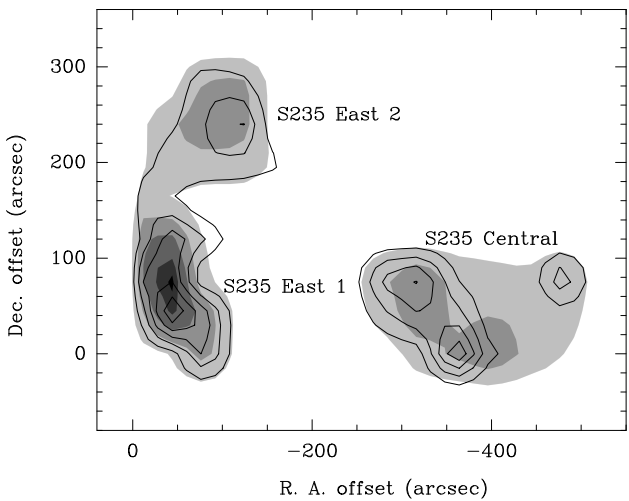}
\caption{Maps of \am(1,1) (grey) and \am(2,2) (contours) emission toward the embedded clusters in the S235 complex. Levels of greyscale correspond to 1.0, 3.0, 5.0, 7.0 and 9.0 K~\kms, contours correspond to 0.25, 0.50, 0.75, 1.00, 1.25 and 1.50 K~\kms. For the reference position, see Fig~\ref{fig:general_view}.}
\label{fig:nh3_12maps}
\end{figure}

\begin{figure*}
\includegraphics[scale=0.65,angle=270]{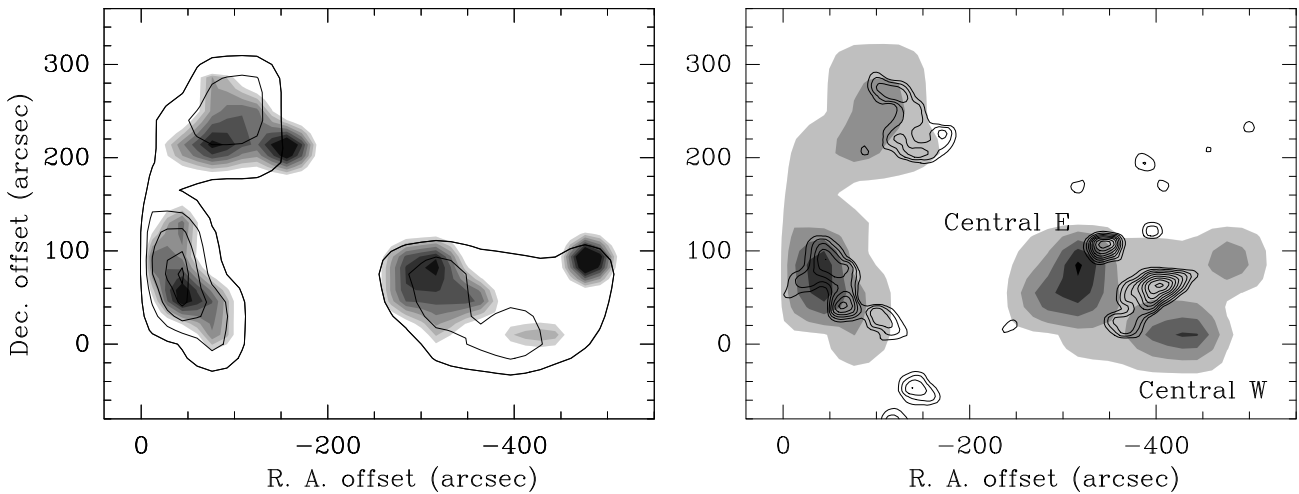}
\caption{Physical conditions in the embedded clusters around S235. Left panel: \tkin\ (grey) and \am(1,1) emission map (contours). The contours correspond to 1.0, 3.0, 5.0, 7.0 and 9.0 K~\kms. Levels of greyscale are given from 10~K to 25~K with a step of 2.5~K. Right panel: Stellar density map taken from~Paper~I (contours) and \nam\ distribution (grey). Contours are the same as in Fig.~\ref{fig:general_view} (right panel). Levels of greyscale are given for $10^{13}$ to $2\cdot 10^{14}$~cm$^{-2}$ with a step of $5\cdot 10^{13}$~cm$^{-2}$. For the reference position, see Fig~\ref{fig:general_view}.}
\label{fig:phys}
\end{figure*}

Maps of \am(1,1) and \am(2,2) emission from \clone, \cltwo, and \clcen\ are shown in Fig.~\ref{fig:nh3_12maps}. The overall structure of the star forming regions is similar to that outlined by the \cs\ map in Fig.~\ref{fig:general_view}. Values of physical parameters in these star forming regions are shown in Fig.~\ref{fig:phys}. Estimates of \tkin\ derived from RT-22 and Effelsberg telescope data are in agreement within the limits of uncertainties.
A spectral map of \am(1,1) toward the S235\,A-B star forming region is shown in Fig.~\ref{fig:spectraAB}. Values of \tkin\ and \gnd\ are also indicated. A complete list of the physical parameters in all these regions is presented in Table~\ref{tab:NH3_res}.

\subsection{\clone}
\label{sec:east1}
\clone\ shows the brightest ammonia emission among \clone, \cltwo, and \clcen. The peak of the \am(1,1) emission is located close to the position with offset~\offsets\,=(--40\arcsec,80\arcsec), while the peak of the \am(2,2) emission is shifted to the south (Fig.~\ref{fig:nh3_12maps}) by 40\arcsec.
Ammonia lines in this clump are optically thin.  The observed line width is $\sim2$~km s$^{-1}$, which is larger than the thermal line width for ammonia emission (0.2--0.7~\kms\ for \tkin\ from 20 to 200~K). So we can infer the presence of non-thermal motions in this clump. Such ammonia line widths are
typical for massive star-forming regions~\citep{jijina,wienen_12}. Radial velocities of the \am(1,1) line centres are in the range from --19 to --21~\kms, being redder than \vlsr\ in the \cltwo\ and \clcen\ clumps.

\subsection{\cltwo}
\label{sec:east2}
Unlike the other two clumps in the dense shell around S235, this region represents only a modest enhancement of ammonia emission. Spectra of ammonia toward the emission peak at \offsets\,=(--120\arcsec,240\arcsec) are shown in Fig.~\ref{ex_spectra}. The spatial distribution of the stellar density is shifted relative to the ammonia \am(1,1) emission and partially coincides with the patch of warm gas at (--160\arcsec,200\arcsec), see Fig.~\ref{fig:phys}. The ammonia lines are optically thin, so lower values of \nam\ and \gnd\ in this clump in comparison with the other clumps are expected because of lower brightness of ammonia emission. In contrast to \clone, there are no ammonia lines centred at --19~\kms. The emission is shifted blueward here. Line widths exceed thermal values, as in the \clone\ clump.

There are regions in this clump, where the \am(1,1) line shows signatures of a non-LTE excitation. One of such regions is found toward the \offsets\,=(--80\arcsec,280\arcsec) position. To describe the anomaly, we follow~\citet{longmore} and define $\alpha_{\rm hyp}$ as a ratio of intensities of an outer left (F$_1=0\rightarrow1$) to an outer right (F$_1=1\rightarrow0$) satellite
\begin{equation}
\alpha_{\rm hyp}=\frac{T_{\rm mb}(F_1=0\rightarrow1)}{T_{\rm mb}(F_1=1\rightarrow0)}.
\end{equation}
Similarly, for the inner satellites
\begin{equation}
\beta_{\rm hyp}=\frac{T_{\rm mb}(F_1=2\rightarrow1)}{T_{\rm mb}(F_1=1\rightarrow2)}
\end{equation}
and for the sum of right and left satellites
\begin{equation}
\gamma_{\rm hyp}=\frac{T_{\rm mb}(F_1=0\rightarrow1)+T_{\rm mb}(F_1=2\rightarrow1)}{T_{\rm mb}(F_1=1\rightarrow0)+T_{\rm mb}(F_1=1\rightarrow2)}.
\end{equation}
In our case, $\beta_{\rm hyp} < 1$ and $\gamma_{\rm hyp} < 1$, namely,
$\beta_{\rm hyp} = 0.52\pm0.22$ and $\gamma_{\rm hyp}=0.66\pm0.20$. Radiative transfer modelling of ammonia line formation in a medium with systematic velocity gradients was performed by \citet{park_01}. He showed that the case of $\gamma_{\rm hyp} < 1$ is realised in a cloud with inward motion. Therefore, it is possible that the \am(1,1) hyperfine intensity anomaly at the offset position \offsets\,=(--80\arcsec,280\arcsec) is created by the line-of-sight velocity gradient in the region emitting in NH$_3$ lines. This case can correspond either to remaining inward motion or deceleration of outflowing gas with distance from the cluster which is typical for expanding flows.

\begin{table*}
\caption{Physical parameters of molecular clumps related to star formation.}
\begin{tabular}{cccccccc}
\hline
 Offset    & N(1,1)        & N(2,2) & N(3,3) & \texit         & \trot          & \tkin        & n(H$_2$)\\
  ($\arcsec$,$\arcsec$)  & (10$^{13}$ cm$^{-2}$)        & (10$^{13}$ cm$^{-2}$)      & (10$^{13}$ cm$^{-2}$)      & (K)         & (K)          & (K)  & (10$^3$ cm$-3$)      \\
 \hline
\multicolumn{8}{l} {{\bf \clone}, ref. pos. $\alpha_{2000} = 05^{\rm h}41^{\rm m}33.\!\!^{\rm s}$8, $\delta_{2000} = +35^{\circ}48'27.\!\!^{\prime\prime}0$}\\
0,160& 2.49 (0.52)   & --          &--  &--           &--            &--            &-- \\
0,120& 4.94 (0.63)   & --          &--  &--           &--            &--            &-- \\
0,80 & 4.17 (0.58)   & 0.76 (0.23) &--  &--           & 18.69 (3.90) & 21.20 (4.42) &-- \\
0,40 & 4.17 (0.43)   & --          &--  &--           &--            &--            &--  \\
--40,160 & 1.12 (0.41)   & --          &--  &--           &--            &--            &--  \\
--40,120 & 8.05 (0.51)   & 1.40 (0.24) & 0.80 (0.63) &10.54 (4.23) & 18.37 (1.93) & 20.73 (2.18) &18\\
 --40, 80  & 19.75 (10.76) & 1.39 (0.93) &--  & 8.36 (2.21) & 16.82 (3.25) & 18.57 (3.58) & 19   \\
 --40, 40  & 15.40 (5.30)  & 2.43 (0.81) & 1.05 (0.38) & 9.24 (1.63) & 24.96 (4.34) & 31.53 (5.48) & 19\\
--80,120 & 1.42 (0.86)   & --          &--  &--           &--            &--            &--  \\
--80,80  & 1.25 (0.45)   & --          &--  &--           &--            &--            &--  \\
--80,40  & 5.85 (0.27)   & 0.89 (0.17) &--  &7.02 (2.74)  & 17.36 (1.76) & 19.31 (1.95) &13\\
 --80, 0  & 4.92 (0.35)   & 0.72 (0.22) &--           &--           & 17.13 (2.72) & 18.99 (3.02) &--\\
\multicolumn{8}{l} {{\bf \cltwo}, ref. pos. $\alpha_{2000} = 05^{\rm h}41^{\rm m}33.\!\!^{\rm s}$8, $\delta_{2000} = +35^{\circ}48'27.\!\!^{\prime\prime}0$}\\
--40, 240  & 2.79 (0.47) & --          & -- &-- &-- &-- &-- \\
 --40, 200 & 3.42 (0.17)   & 0.43 (0.11) &--  &-- & 16.06 (1.91) & 17.56 (2.09) &--\\
 --80, 280  & 4.36 (0.22)   & 0.48 (0.09) &-- &-- & 15.39 (1.34) & 16.69 (1.45) &--\\
 --80, 240  & 5.17 (0.65)  & 0.66 (0.20) &-- &-- & 16.17 (2.83) & 17.69 (3.10) &--\\
--80, 200  & 6.46 (4.92)  & 1.81 (1.45) &-- &3.91 (0.69) & 23.31 (9.64) & 28.56 (11.81) &3\\
 --120, 200 & 3.45 (0.34)  & 0.44 (0.15) &-- &--  & 16.15 (2.67) & 17.68 (2.92) &--\\
 --120, 240 & 5.64 (0.37)  & 0.97 (0.15) & 0.23 (0.20) &-- & 18.42 (1.30) & 20.81 (1.47) &--\\
--120, 280 & 5.08 (0.40) & --          &--  &--           &--            &--            &--  \\
 --160, 200 & 0.94 (0.12) & 0.35 (0.10) & 0.11 (0.04) &-- & 27.44 (2.65) & 36.45 (3.52) &--\\
\multicolumn{8}{l} {{\bf \clcen}, ref. pos. $\alpha_{2000} = 05^{\rm h}41^{\rm m}33.\!\!^{\rm s}$8, $\delta_{2000} = +35^{\circ}48'27.\!\!^{\prime\prime}0$}\\
 --280, 80 & 3.48 (0.28) & 0.54 (0.19) & 0.17 (0.12) &-- &19.56 (2.40) &22.48 (2.75) &--\\
--280, 40 & 12.37 (9.29) & 2.43 (2.37) &--& 3.30 (0.35) & 19.40 (6.80) & 22.25 (7.79) & 1\\
 --320, 80 & 20.80 (7.20) & 2.85 (1.43) &--& 4.35 (0.40) & 22.94 (4.26) & 27.92 (5.19) & 4 \\
 --320, 40 & 16.10 (2.30)  & 1.50 (0.31) & 0.38 (0.23) & 5.14 (0.24) & 18.97 (1.19) & 21.61 (1.34) & 6 \\
--320, 0 & 2.56 (0.93) & 0.13 (0.20) &-- &-- & 13.49 (10.67) &14.32 (11.34) &-- \\
--360, 80 & 2.44 (0.60) & --          &--  &-- &--            &--            &--\\
 --360, 40      & 3.59 (0.36) & 0.62 (0.22) &--  &-- & 18.33 (3.79) & 23.65 (4.35) &--\\
--360, 0  & 3.55 (0.89) & --          &--  &-- &--            &--            &--\\
--400, 80 & 2.56 (0.75) & --          &--  &-- &--            &--            &--\\
--400, 40 & 4.12 (0.37) & --          &--  &-- &--            &--            &--\\
--400, 0  & 11.96 (9.38)& 0.76 (0.95) &--  &4.42 (1.01) & 12.72 (3.16) & 13.41 (3.33) & 5\\
--440, 80 & 1.89 (0.41) & --          &--  &-- &--            &--            &--\\
--440, 40 & 2.95 (0.45) & --          &--  &-- &--            &--            &--\\
--440, 0  & 8.87 (5.84) & 8.29 (7.23) &--  &3.52 (0.38) & 14.40 (3.37) & 15.44 (3.62) & 2\\
 --480, 80 & 8.71 (3.20)   & 1.84 (0.70) & 0.52 (0.49) & 5.14 (0.66) & 30.27 (7.46) & 43.04 (10.62)& 5 \\
 \multicolumn{8}{l} {{\bf Primordial gas}, ref. pos. $\alpha_{2000} = 05^{\rm h}41^{\rm m}33.\!\!^{\rm s}$8, $\delta_{2000} = +35^{\circ}48'27.\!\!^{\prime\prime}0$}\\
   --600, --150 & 1.52  (0.09)  & 0.45 (0.15) & 0.07 (0.03) &--  & 22.66 (2.18) & 27.44 (2.65) & -- \\
 --600,   0     & 1.66  (0.23)  & 0.38 (0.14) & --          & -- & 20.75 (5.48) & 24.33 (6.42) & -- \\
--200, 400 & 0.49 (0.11) & --          &--  &--           &--            &--            &--  \\

\multicolumn{8}{l} {{\bf {\bf S235\,A-B}}, ref. pos. $\alpha_{2000} = 05^{\rm h}40^{\rm m}53.\!\!^{\rm s}$4, $\delta_{2000} = +35^{\circ}41'48.\!\!^{\prime\prime}0$}\\
 0,40    & 1.55  (0.84)  & --          &--  &-- &--            &--            &--\\
 0,20    & 19.26 (16.64) & 3.26 (2.58) & 4.06 (4.28) & 4.92 (1.31)  & 26.06 (13.28) & 33.63(17.14) & 5 \\
--40,0   & 0.51 (0.11)   & --          &--  &-- &--            &--            &--\\
    0,0  & 21.07 (12.76) & 5.83 (3.53) & 2.71 (1.62) & 10.18 (3.22) & 37.68 (17.27) & 66.09 (30.40) & 16 \\
40,0     & 3.50 (0.49)   & 0.72 (0.42) & --          &--  & 19.80 (7.80)  & 22.86 (9.00) &--\\
20,--20  & 22.31 (23.34) & 2.35 (2.24) & 2.70 (2.64) & 13.72 (7.80) & 19.66 (7.56)  & 22.64 (8.71) & 51\\
 0,--20  & 18.60 (0.70)  & 2.22 (0.41) & 1.79 (0.84) & -- & 21.86 (1.49)  & 26.12 (1.78) &--\\
--20,--20& 3.87  (0.80)  & 1.91 (1.07) & 0.80 (0.74) & -- & 33.26 (12.70) & 51.41 (19.63)&--\\
 0,--40  & 13.78 (0.67)  & 2.17 (0.60) & 2.50 (2.05) & -- & 18.01 (1.92)  & 20.22 (2.15) &--\\
 0,--60  & 19.55 (0.83)  & 1.12 (0.34) & 0.75 (0.55) & -- & 14.73 (1.30)  & 15.85 (1.40) &--\\
--20,--60& 10.59 (0.83)  & 1.56 (0.56) & 1.28 (0.79) & -- & 22.50 (2.90)  & 27.17 (3.51) &--\\
20,--60  & 14.05 (0.81)  & 2.09 (0.95) & 1.63 (0.70) & -- & 25.45 (2.78)  & 32.36 (3.54) &--\\
 0,--80  & 26.43 (11.61) & 2.30 (1.11) & 1.58 (1.01) & 9.49 (2.05)  & 18.38(3.08)   & 20.76(3.48) & 23 \\
 0,--100 & 11.41 (0.83)  & 2.73 (1.23) & --          & -- & 21.38 (6.14)  & 25.35 (7.28) &--\\
 0,--120 & 1.39  (0.77)  & 0.45 (0.25) & --          & -- & 25.27 (7.10)  & 32.10 (8.40) &--\\
\hline
\end{tabular}
\label{tab:NH3_res}
\end{table*}

\subsection{\clcen}
\label{sec:central}

Ammonia emission in this clump has several peaks. The spatial distribution of \am(1,1) is different from that of \am(2,2) as shown in Fig.~\ref{fig:nh3_12maps}. In some positions the optical depth of ammonia is close to unity in contrast to lower values encountered in \clone\ and \cltwo\ (see Table~\ref{tab:obs_res}). Spectra toward the strongest peak of \am(1,1) emission at \offsets\,=(--320\arcsec,40\arcsec) are shown in Fig.~\ref{ex_spectra}. Radial velocities obtained from Gaussian fits range from --19~\kms\ to --22~\kms. The scatter in radial velocities between different parts of \clcen\ is higher than that in \clone\ and \cltwo. Widths of ammonia lines are also super-thermal here.

A hyperfine intensity anomaly is possible around the position with the offset \offsets=(--320\arcsec,40\arcsec) as seen in Fig.~\ref{ex_spectra}. The significance of the deviation is about 1.5 sigma. Here we have $\alpha_{\rm hyp}=1.33\pm0.21$ and $\beta_{\rm hyp}=0.84\pm0.11$, that is, the case of $\alpha_{\rm hyp} > 1$ and $\beta_{\rm hyp} < 1$, which was explained by \citet{matsakis} and \citet{stutzki_hyp}. According to these studies, several small (in comparison to the beam size of the Effelsberg telescope at 23~GHz, i.e. 40\arcsec) dense clumps contribute to the observed \am(1,1) line. Individual clumps should have line widths of about 0.3--0.6~\kms\ which leads to selective trapping in the hyperfine components of the far infrared ($J,K) = (2,1)\rightarrow(1,1)$ transition via Doppler shift.

\subsection{Quiescent primordial gas}
\label{sec:primord}

We have obtained ammonia spectra toward three positions where primordial gas is supposed to be located. It can be seen from Fig.~\ref{fig:nh3_12maps} and Table~\ref{tab:obs_res} that the brightness of \am(1,1) and \am(2,2) lines is significantly lower there in comparison with ammonia emission from the dense clumps. The spectra toward the (--600\arcsec,--150\arcsec) position are shown in Fig.~\ref{ex_spectra}. Line widths exceed thermal values, so non-thermal motions exist in the primordial gas, too. Radial velocities of line centres exceed --18~\kms, that is, lines from the dense clumps are shifted blueward relative to those from primordial gas. The primordial gas is quite warm, with \tkin\ of about 27~K. \nam\ is about 2$\cdot 10^{13}$~cm$^{-2}$ there.

\subsection{S235\,A-B}

\begin{figure}
\includegraphics[scale=0.43,angle=0]{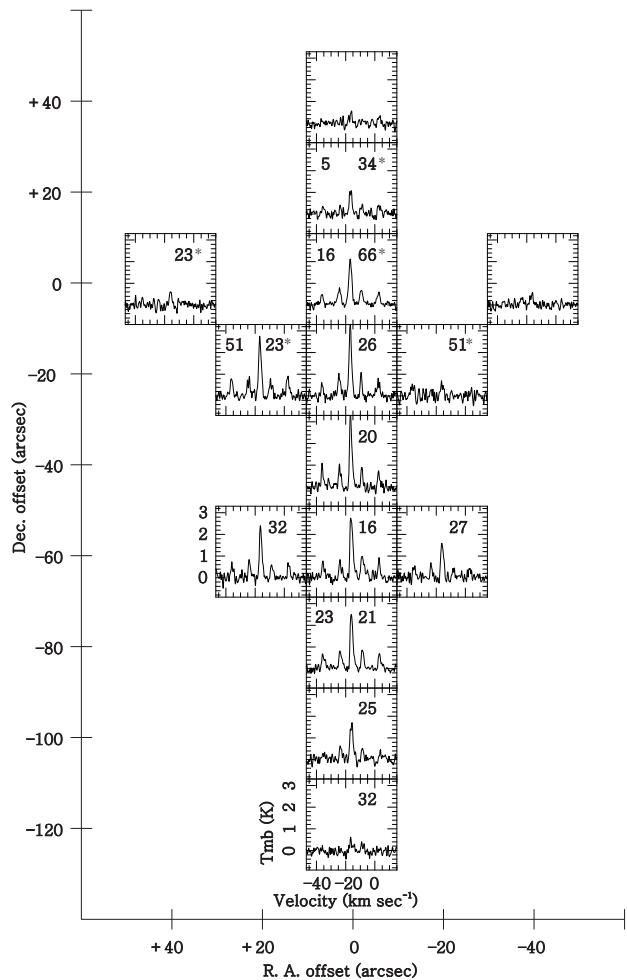}
\caption{Spectra of \am(1,1) emission toward the S235\,A-B region. Offsets are given with respect to $\alpha_{2000} = 05^{\rm h}40^{\rm m}53.\!\!^{\rm s}$4 and $\delta_{2000} = +35^{\circ}41'48.\!\!^{\prime\prime}$0. Values of \tkin\ (in K) and gas number density (in units of 10$^3$ cm$^-3$) are given in the upper right and upper left hand sides of the \am(1,1) main component, respectively. Asterisks denote values with uncertainties exceeding 40~per cent.}
\label{fig:spectraAB}
\end{figure}

The brightest peak of ammonia emission in the entire S235 area is found toward the star forming region S235\,A-B. Spectra of \am(1,1) emission are shown in Fig.~\ref{fig:spectraAB}. The \hii\ region \cla\ itself is located to the north from (0\arcsec,0\arcsec) position at Fig.~\ref{fig:spectraAB} (reference position is $\alpha_{2000} = 05^{\rm h}40^{\rm m}53.\!\!^{\rm s}$4 and $\delta_{2000} = +35^{\circ}41'48.\!\!^{\prime\prime}$0), with offset \offsets=(0\arcsec,34\arcsec) (see also Fig.~\ref{fig:general_view}). The vicinity of the \cla\ and \clb\ nebulae is only partially covered by our data, mainly consisting of spectra along the north-south direction. It allows investigating the gradient of physical parameters along the major axis of the region. Ammonia emission in S235\,A-B is optically thin (Fig.~\ref{tab:obs_res}). Line widths exceed thermal values. Radial velocities are redshifted relative to those in \clone, \cltwo, \clcen\ and are similar to velocities of the primordial gas.

\section{Physical structure of the star forming regions}
\label{SEC:DISC}

\subsection{Summary of densities and temperatures}

The ammonia emission in the entire S235 area is very non-uniform. It consists of several bright clumps that generally coincide with peaks in stellar density and emission of other molecular tracers.

In those locations within the northern clumps (\clone, \cltwo\, and \clcen), where physical parameters can be determined, ammonia column density varies from $\sim10^{13}$~cm$^{-2}$ to $2\times10^{14}$~cm$^{-2}$. The value of $N_{{\rm NH}_3}$ toward the primordial gas locations is about $10^{13}$~cm$^{-2}$ (solid circles in Fig.~\ref{fig:general_view}). Temperatures over the same area vary from 13 to 43\,K. Temperature distribution apparently do not follow any simple pattern, with ``warm'' and ``cold'' spots being generally intermixed with each other. In \clcen\ and \cltwo\ clumps high temperature spots are located at their respective inner boundaries, in relative isolation. While this arrangement may be indicative of gas heating by UV~emission from S235, heating by internal sources is also possible. For example, the warm spot in \cltwo\ at (--160\arcsec,200\arcsec) is situated only $\sim1$\arcmin\ off a bright millimetre continuum source \citep{klein_05}, while the warm spot in \clone\ at (--40\arcsec,40\arcsec) may have appeared due to illumination from IRAS 05382+3547. Also, abrupt \tkin\ changes observed in the studied clumps may be an indication of unresolved embedded heating sources. Were the HII region entirely responsible for gas heating, we would expect warm regions to be smoother and more extended.

The brightest \am(1,1) emission in the three clumps, embedded in the molecular shell around S235, comes from gas in \clone. This gas is in average denser and colder than gas in the other clumps of the shell. An ammonia column density \nam $\sim2\times 10^{14}$~cm$^{-2}$ is found toward the \am(1,1) peak in \clone. This high column density corresponds to low \tkin\ (about 19~K) and high gas volume density (about $2\times 10^{4}$~cm$^{-3}$ toward the centre of the clump).

There is no notable emission and column density peak in \cltwo. The \nam\ distribution is smooth with values of about a few times $10^{13}$~cm$^{-2}$ at nearly all the observed locations. Unlike \clone, the highest column density in \cltwo\ corresponds to relatively high temperature, but in other locations within the clump \tkin\ values are low. The gas volume density is only determined at the location of the highest column density and is nearly an order of magnitude lower than in \clone.

\clcen\ contains at least two dense subclumps (Fig.~\ref{fig:phys}) with column densities of about $10^{14}$~cm$^{-2}$ (Central~W) and $2\times10^{14}$~cm$^{-2}$ (Central~E). The greatest column density in \clcen\ exceeds that in \clone\ despite emission here being fainter as effects of the optical depth could be important in \clcen\ ($\tau(1,1) \sim 1$ at the emission peaks). Even higher values of \nam\ can be expected in the densest part of \clcen\ because of effects of self-absorption.

The volume density in the dense subclumps is $4-6\times 10^{3}$~cm$^{-3}$ and an order of magnitude lower ($\sim5\times 10^{2}$~cm$^{-3}$) in the region between them (see Table~\ref{tab:rt22_res}). We note again that the actual density can be higher because of the optical depth effects. As in \cltwo, the location with the highest \nam\ in \clcen\ is quite warm, with \tkin\ being above 25~K. The lowest temperatures in the shell are found in the southern part of \clcen, where the peak of CO emission is observed.

The highest values of \tkin\ and \gnd\ among all the star forming regions around S235 are found in S235\,A-B (see Table~\ref{tab:NH3_res} and Fig.~\ref{fig:spectraAB}). The value of \tkin\ decreases from more than 50~K at the water maser position (0, 0) to 16~K at the position 1\arcmin\ to the south, and then rises again to 30~K at the position 2\arcmin\ to the south. The value of \trot\,=~37~K toward the water maser is consistent with that of the CH$_3$CN(5--4) transition \citep{felli_s235ab}. The maximum gas density is found in about 20\arcsec\ to the east from the centre of the S235\,B nebula (see Fig.~\ref{fig:general_view} for location of the S235\,B in S235\,A-B star forming region, and Fig.~\ref{fig:spectraAB} for the spectra). Here \gnd\, is larger by a factor of two than the density in the densest part of \clone\ ($5.1\times10^4$~cm$^{-3}$ and $1.9\times10^4$~cm$^{-3}$, respectively). Gradients of \tkin, \gnd\ and \nam\ in the north-south direction, that we found, reflect the large scale structure of the star forming region to the south of \clb. Gas temperature rises as we move southward from S235\,B to the region with bright HCO$^+$ and $^{13}$CO emission (P. Boley et al., in preparation). Young star cluster and \hii\ region S235\,C are located there (beyond our maps).

\subsection{Radial velocities of the molecular gas}
\label{sec:radvel}

In Paper~I, discrimination between the quiescent undisturbed gas and gas which has been compressed by the shock from the expanding \hii\ region was made on the basis of gas kinematics reflected in \cs\ and \co\ emission. Specifically, the ``red'' \vlsr\ range (from --15~\kms\ to --18~\kms) represents quiescent undisturbed primordial gas. The range from --18~\kms\ to --21~\kms\ (hereafter the ``central'' range) corresponds to gas compressed by the shock from expanding S235 \hii\ region. The ``blue'' range from --21~\kms\ to --25~\kms\ presumably corresponds to gas expulsion from the embedded young star clusters driven by combined effect of the cluster stars.

\begin{figure*}
\includegraphics[width=0.7\textwidth]{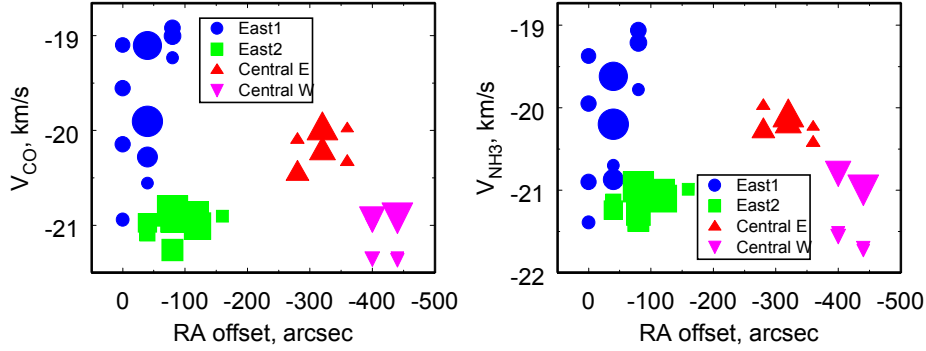}
\caption{Line centre velocities for CO (left panel) and NH$_3$ (right panel) emission as a function of offset in right ascension (at all declinations). Different symbols correspond to various gas clumps as indicated in the legend. A symbol size corresponds to the column density. The colour version of this figure is available in the electronic edition of the journal.}
\label{kinem01}
\end{figure*}

Ammonia emission supports this picture. Radial velocities of \am\ lines in the dense molecular clumps \clone, \cltwo, and \clcen\ mostly belong to the ``central'' range, while \vlsr\ values from the primordial gas are about --17~\kms, that is, belong to the ``red'' velocity range. These velocities are consistent with \vlsr\ of \co\ and \cs\ emission lines which have higher optical depths. Radial velocities of ammonia lines toward the star forming region S235\,A-B are similar to velocities of the primordial gas and belong to the ``red'' range. This is consistent with our suggestion from Paper~I that star formation in S235\,A-\,B is independent from the expansion of S235. In summary, the overall  kinematic structure of the S235 environment, suggested in Paper~I, is confirmed by optically thin ammonia emission. This is illustrated in Fig.~\ref{kinem01} where we show $^{13}$CO and ammonia velocities determined from Gaussian fits as a function of offset in RA (note that in both panels we only show data for positions where ammonia emission is present).  The two subclumps in \clcen\ apparently have distinct average velocities, namely, about --20 \kms\, for \clcen~E and about --21 \kms\, for \clcen~W.

\begin{figure}
\includegraphics[scale=0.3,angle=270]{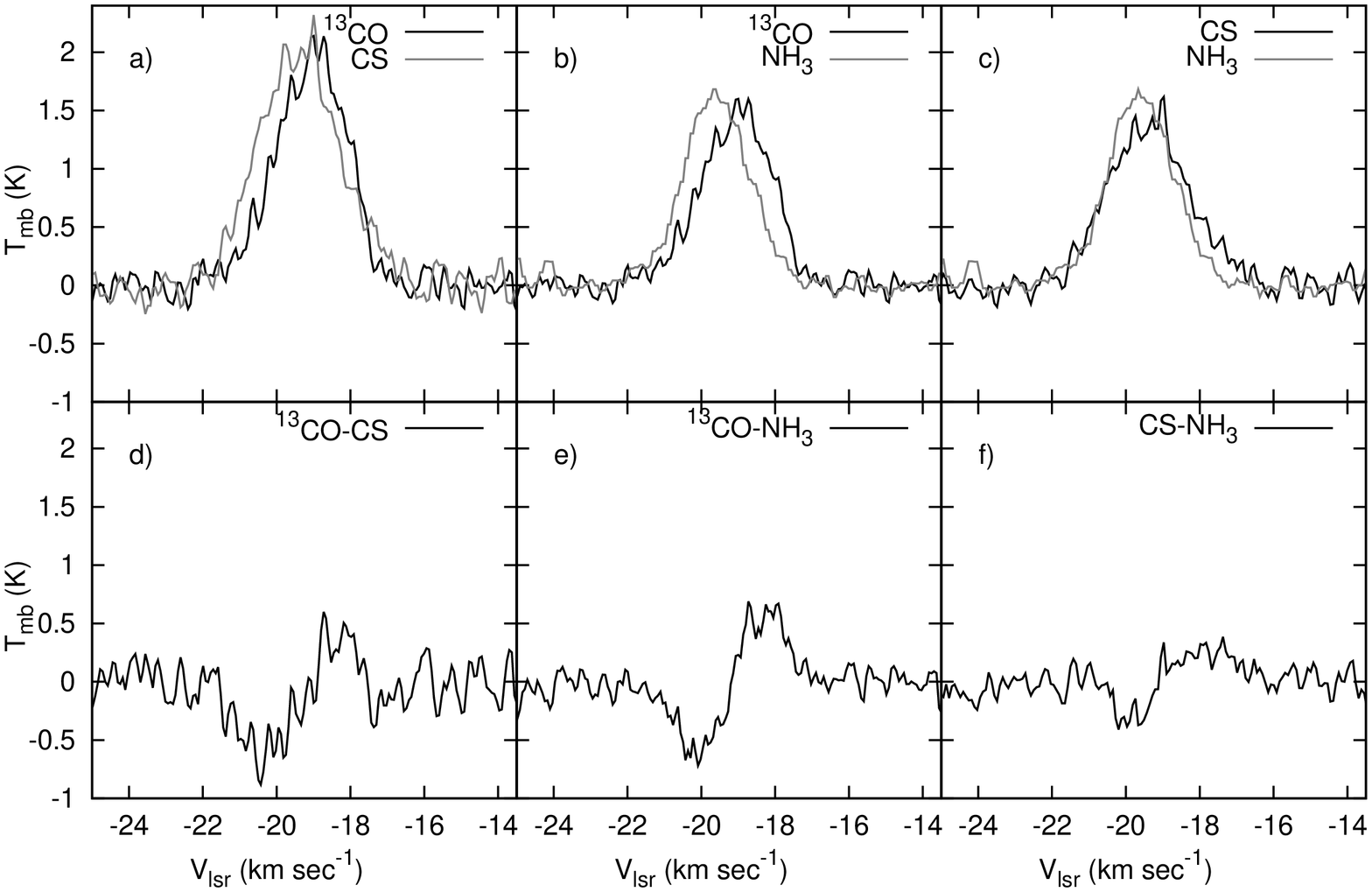}
\caption{Scaled \cs, \co, and \am(1,1) spectra toward the (--40\arcsec,40\arcsec) position (top row) and the difference between the corresponding spectra (bottom row). a) \co$\cdot0.09$ and \cs, b) \co$\cdot0.06$ and \am(1,1), c) \cs$\cdot0.7$ and \am(1,1), d) \co$\cdot0.09-$\cs, e) \co$\cdot0.06-$\am(1,1), f) \cs$\cdot0.7-$\am(1,1). For the reference position, see Fig~\ref{fig:general_view}.}
\label{fig:velshift}
\end{figure}

Apart from the three velocity components, yet another kinematic feature found in Paper~I is a velocity shift between \co\ and \cs\ lines in \clone\ (Fig.~\ref{fig:velshift}a), reaching about 2\,\kms\ in line wings. As both CS and NH$_3$ trace dense gas, we may expect a similar velocity difference between \co\ and ammonia lines, and it indeed exists. An example of the shift of \am(1,1) and \co\ lines toward the (--40\arcsec,40\arcsec) offset position is presented in Fig.~\ref{fig:velshift}a,b along with the \co\ versus \cs\ shift. Both \co\,--\cs\ velocity shift and \co\,--\am(1,1) velocity shift presumably indicate the relative motion of a dense clump in \clone\ and its outer envelope. \am(1,1) and \cs\ line profiles are also slightly shifted relative to each other, but in this case the shift is seen mostly in the red wings of the lines (Fig.~\ref{fig:velshift}c). To show the shifts more clearly, in panels Fig.~\ref{fig:velshift}d, e, and f we display a channel-by-channel subtraction of \co\,--\cs\, and ammonia spectra, scaled as indicated in the caption. The shifts \co\ versus \cs\ and \co\ versus \am\ are smaller than the widths of the \co, \am, and \cs\ lines themselves and are comparable to the thermal width for molecular gas at the temperature of 20--30\,K (0.3--0.4\,\kms).

While the \co--NH$_3$ line shift is most clearly seen toward the (--40\arcsec,40\arcsec) position, it is not a unique feature of this location. \co\ and ammonia lines are somewhat displaced relative to each other at other locations as well. To quantify this shift, we subtracted ammonia line centre velocities (main component) from \co\ line centre velocities as determined from gaussian fits. A shift like this has been analysed previously in several star forming regions. In particular, \cite{walsh_04} looked for motions of dense cores (in the N$_2$H$^+$(1-0) line) with respect to their more rarefied envelopes (using \co\ and C$^{18}$O(1-0)) toward a sample of low-mass star-forming cores. Very small velocity differences (less or about 0.1~\kms) were found, so \cite{walsh_04} concluded that isolated low-mass cores generally do not move ballistically relative to their surrounding envelopes. Similar analysis for regions of massive star formation was performed by \cite{wienen_12}. They discussed relative motions in several high-mass cold clumps based on observed \co\ and NH$_3$ lines and mentioned some trends that can be compared to our results.

\begin{figure*}
\includegraphics[width=0.7\textwidth]{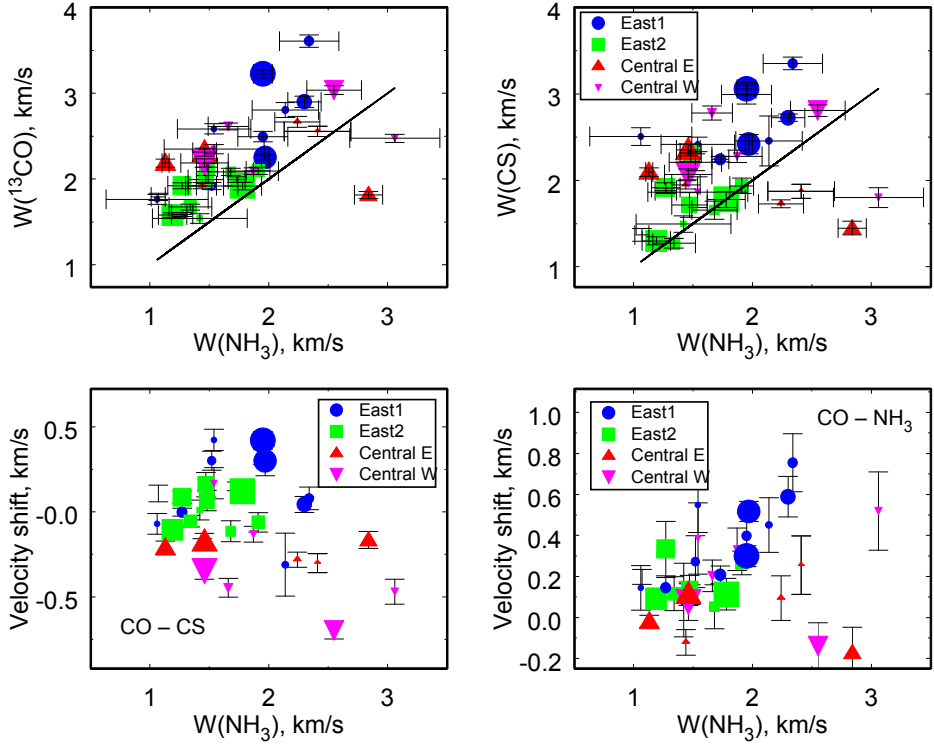}
\caption{{\em Top row}: widths of the $^{13}$CO lines and CS lines compared to those of ammonia (1,1) lines. Solid line corresponds to equality of both widths. {\em Bottom row}: Velocity shift between a low density tracer (CO) and higher density tracers (CS and NH$_3$) as functions of the ammonia (1,1) line width. Different symbols correspond to various gas clumps as indicated in the legend. A symbol size corresponds to the column density. The colour version of this figure is available in the electronic edition of the journal.}
\label{kinem02}
\end{figure*}

First, \cite{wienen_12} found that $^{13}$CO linewidths are as a rule broader than widths of NH$_3$ (1,1) lines. This holds in our study as well, both for CO and CS lines, even though at a smaller scale, as clumps that we have observed are less dense and more compact. Ammonia (1,1) linewidths range from 1 to 3\,\kms, while $^{13}$CO and CS linewidths are in average 20--30\% larger and extend to 3.5\,\kms\ (Fig.~\ref{kinem02}, top row). Among positions with high column density, the only one with a wider ammonia line and narrower CO and CS lines is the position (--280\arcsec,40\arcsec) in \clcen~E. Apart from this, all the gas clumps seemingly follow the same pattern in ``width vs width'' plots.

This is not true in the case of line velocity differences. In \clone\ lines of a low density tracer (\co) are shifted relative to higher density tracers (\cs\ and NH$_3$), as we noted in Paper~I. The maximum shift between \co\ and \cs\ line profiles is about 0.5 km/s which is a bit higher than the thermal velocity and significantly lower than the line widths (Fig.~\ref{kinem02}, bottom left, filled circles). With respect to \co\--\am, some regularity is seen in \clone: larger ammonia line widths correspond to higher velocity differences between \co\ and ammonia (Fig.~\ref{kinem02}, bottom right, filled circles).

In \cltwo\ and \clcen~E there is almost no difference between $^{13}$CO and CS line velocities, and the difference between $^{13}$CO and NH$_3$ line velocities is much smaller than in \clone\ (but also positive). In \clcen~W the difference between CO and ammonia velocities is close to zero, and CO~--~CS difference is notably negative.

The analysis performed by \cite{wienen_12} has shown that supersonic velocity differences are typical of high-mass clumps, unlike low-mass star-forming regions, where velocity differences between low- and high-density tracers are mostly subsonic. Our data show a mixed picture. While the CO~--~CS difference is at best mildly supersonic (in \clone\ and \clcen~W), CO~--~NH$_3$ differences exceed the sound speed more significantly, reaching almost 1\,\kms\ in \clone. The important difference between our results and \cite{wienen_12} results is that CO~--~NH$_3$ differences that we measure are positive at almost all locations where ammonia emission is observed. A possible explanation is that \cite{wienen_12} studied large-scale regions. In this case, the velocity differences can be related to chaotic motions of individual dense clumps within a low-density medium. We study individual clumps themselves, so our results are, probably, better explained in the ``core-to-envelope'' paradigm. More specifically, as we relate gas clumps to star clusters, they may represent gas expelled from the clusters, with different lines tracing denser and less dense parts of an expanding envelope.

In Paper~I we argued that \clone\ is the youngest cluster in the northern part of the S235 region. The ammonia data are in qualitative agreement with this suggestion. The ammonia lines being blue-shifted with respect to CO indicates the presence of a significant gas velocity gradient along the line of sight. An outflow in the region is not well developed and contains considerable amounts of weakly processed gas. This gas is mostly in molecular form and displays considerable variability in the outflow kinematics with the distance from the cluster. Molecular gas is observed in projection onto the cluster, so the difference in velocities of molecular lines which trace regions with different kinematics is well pronounced. It arises both due to chemo-dynamical processing of the gas in a propagating shock and ionization front \citep[see, e.g.][]{Kirsanovaetal2009} and also due to line excitation details. Ammonia emission arises in the dense gas that has been swept up from the cluster by the combined action of young stars and now moves toward the observer. CO emission forms in less dense regions and attains significant optical depth much closer to the observer. The more red-shifted CO velocity component may trace the deceleration of the expanding outflow or leftovers of the parent gas clump which are still moving toward the \clone\ cluster.

The absence of a similar velocity shift toward \cltwo\ and \clcen~E is also consistent with our suggestion that clusters embedded in these clumps are actively losing leftover molecular gas that shows up observationally as \co\ and \cs\ emission in the ``blue'' range. At this later evolutionary stage any initial kinematical structure, including velocity shifts between various molecular tracers, has probably been destroyed. It must be noted that while both these clumps possess CO and CS emission in the blue range and are somewhat displaced relative to corresponding clusters, otherwise they are kinematically different. The \cltwo\ clump is very quiescent, with small linewidths that do not differ much from one direction to another. On the other hand, \clcen~E is more disturbed showing both wide and narrow lines, sometimes in adjacent directions.

Kinematic properties of \clcen~W are harder to interpret as it differs from the other clumps in a couple of aspects. First, this is the coldest clump among those covered with our observations. Also, in the other clumps CS and ammonia line positions are generally consistent with each other, in the sense that CO~--~CS and CO~--~NH$_3$ velocity differences are either both close to zero (in \cltwo\ and \clcen~E) or both positive (in \clone). In \clcen~W velocities of CO and NH$_3$ are similar, but CS emission is shifted significantly blueward relative to CO emission. This difference does not fit into our simple picture and craves for more data.

\section{Star formation triggering around S235}
\label{sec:trig}

\subsection{Sizes and triggering timescales}

In Paper~I we proposed that both the ``collect-and-collapse" (C\&C) process and compression of pre-existing clumps could be responsible for the star formation in this region. Here we re-consider this suggestion using  S235 parameters derived from our ammonia observations. Specifically, we compare our results with predictions of a theoretical model by \citet{whitworth_94} to investigate the history of star formation around S235 and to determine main characteristics of a C\&C process: time, when fragmentation of a dense shell starts (\tfrag), the radius of the shell (\rfrag) at this moment, and mass (\mfrag) of the fragments. We also calculate the radius of an initial Str\"omgren sphere, \iss,~\citep{dyson_williams_80}. Then we relate these values to parameters of S235 and \cla\ in order to estimate ages of the \hii\ regions. Given the clumpiness of matter, surrounding both regions, the model of \citet{whitworth_94} should be applied with care, and our conclusions can only be considered as approximate.

To use the relations from \citet{whitworth_94}, one needs to specify some parameters. To calculate the speed of sound in the undisturbed medium, we adopt 25~K as a typical temperature for S235 and 34~K as a typical temperature for \cla. The gas number density in the undisturbed medium is supposedly equal to the density of primordial gas for S235 and to the density of surrounding gas for \cla. Several values of $n_0$ are considered for S235. The lowest one, 500\,cm$^{-3}$, is an average large-scale value measured with the RT-22 (see Table~\ref{tab:rt22_res}). The observationally motivated upper limit is taken to be $3\cdot10^3$\,cm$^{-3}$ which is a reasonable value for the region of S235 according to~\citet{evans_81} (their ``--\,17\kms" cloud). We also consider a higher value of $7\cdot10^3$\,cm$^{-3}$ to find the density at which the estimated \hii\ region age becomes comparable to \tfrag. The flux of hydrogen-ionising photons ($S_{\rm ion}$) was taken from~\citet{sternberg_03} in accordance with the spectral types of ionising stars which are O9.5 in S235~\citep{georg_73} and B0.5 in \cla~\citep{thompson_83}.

Distance estimates $D$ to S235 vary from 1.6 to 2.0~kpc. We adopted an intermediate value $1.8$\,kpc\ \citep{evans_81} for our estimates. The angular size $\varphi$ of S235 is 380\arcsec\ $\times$ 450\arcsec\ in accordance with radio continuum measurements \citep{isr_felli}. Thus, the linear radius of S235 is constrained by values from 1.5\,pc ($D=1.6$\,kpc, $\varphi=380$\arcsec) to 2.2\,pc ($D=2.0$\,kpc, $\varphi=450$\arcsec). We adopt 2\,pc as a representative value for S235. The angular size of \cla\ is 20\arcsec\ \citep{isr_felli}, so its linear radius is between 0.08 and 0.10~pc. All the adopted parameters for S235 and \cla\ are summarised in Table~\ref{tab:comparison}. Results of a comparison with calculations based on~\citet{whitworth_94} are given in Table~\ref{tab:candcres}. In the last row we present an \hii\ region age, estimated using Eq.~4 from \cite{whitworth_94}.

\begin{table}
\caption{Parameters of S235 and \cla\ used for calculation of characteristics of the triggering process.}
\begin{tabular}{lcc}
\hline
Parameter     & S235 & \cla\                   \\
\hline
$S_{\rm ion}$, sec$^{-1}$ & 10$^{48.26}$ & 10$^{48.02}$ \\
\tkin, K & 25 & 34 \\
$n_0$, cm$^{-3}$ & $5\cdot10^2-3\cdot 10^3$ & $5\cdot 10^3$\\
\hline
Radius, pc  & 2.0$^{2.2}_{1.5}$& 0.09$^{0.10}_{0.08}$\\
\hline
\end{tabular}
\label{tab:comparison}
\end{table}

The fragmentation radius  \rfrag\ coincides with the actual linear radius of the S235 \hii\ region within a factor of 2 only for $n_0$ of $3\cdot10^3$\,cm$^{-3}$ or greater. So, it is possible that triggering of star formation in \clone, \cltwo, and \clcen\ by a C\&C process took place in the recent past if the density in the surrounding medium is close to the upper limit of the range that is indicated by our observations (typically about $3-5 \cdot 10^3$\,cm$^{-3}$ toward non-peak positions). In this case, star-forming clumps should have masses of about 50-100\,$M_\odot$.

It is interesting that the projected distance from the exciting star of S235, BD~+35$^{\circ}$1201~\citep{georg_73}, to \cla\ and \clb\ is close to \rfrag\ for $n_0\sim3\cdot10^3$\,cm$^{-3}$. However, as already mentioned, the star forming regions S235\,A-B are located outside the common dense envelope that contains \clone, \cltwo, and \clcen. The outer boundary of the S235 \hii\ region is about two times closer to \clone\ and \cltwo\ than to \cla\ (see Fig.\,\ref{fig:general_view}). A bright bubble of the S235 photon dominated region on {\it Spitzer}-IRAC images does not have a common border with \cla. Therefore the formation of \cla\ and \clb\ exciting stars could not have been triggered by the expansion of S235. This conclusion is confirmed by \citet{dewangan_11}. They found that YSOs in the star forming region S235\,A-B are about 10-20 times older than YSOs in \clone\ and \cltwo. Ages of embedded clusters around S235 and in S235\,A-B were also estimated by \citet{camargo_11}  to be $3\pm2$\,Myrs. However, the low accuracy of these results precludes them from being used to elucidate the star formation history in the area.

\begin{table}
\caption{Estimated characteristics related to a collect-and-collapse process for the parameters derived for S235 and \cla.}
\begin{tabular}{lccccc}
\hline
Value       & \multicolumn{4}{c}{S235}& \cla\\
\hline
$n_0$       & $5\cdot10^2$ & $10^3$& $3\cdot10^3$& $7\cdot10^3$& $5\cdot10^3$ \\
\hline
\iss, pc   & 0.7         & 0.4  & 0.2        & 0.1        &0.1 \\
\rfrag, pc & 9          & 6  & 3         & 2         &3\\
\tfrag, Myr& 3          & 2  & 2         &  1        & 1\\
\mfrag, $M_\odot$ &200     & 150   & 90         &  60        & 140\\
Age, Myr            & 0.3     & 0.4   & 0.6         &  1        & 0.004\\
\hline
\end{tabular}
\label{tab:candcres}
\end{table}

A central star in \cla\ has a spectral type close to that of the star in S235. Its ionising power is only about 2 times smaller. Therefore the much smaller size of \cla\ suggests that it is much younger. Note, however, that \cla\ expands in a medium with increasing density (see Table~\ref{tab:NH3_res} and Fig.~\ref{fig:general_view}). Specifically, in \cla\ \gnd\ is about $5\cdot 10^3$~cm$^{-3}$ toward the position with offset \offsets=(0\arcsec,20\arcsec) (see Fig.~\ref{fig:spectraAB}) and $\sim 16\cdot 10^3$~cm$^{-3}$ toward the position with offset \offsets=(0\arcsec,0\arcsec). Remember that the \hii\ region \cla\ itself is located at (0\arcsec,34\arcsec). Given this, it is hard to estimate an actual age of \cla. If we take the lower value of \gnd\,$\sim 5\cdot 10^3$~cm$^{-3}$ and assume a uniform density distribution, we can estimate the minimum possible age of \cla, using formulae from~\citet{whitworth_94}. The adopted physical size of the \cla\ \hii\ region\ is close to \iss\ and significantly smaller than \rfrag\ (see Table~\ref{tab:candcres}). If we again use Eq. (4) from \cite{whitworth_94}, we obtain an age of only 4000 years for \cla. The upper limit of \gnd, quoted above, increases this value by a factor of 2. Anyway, the estimated age of \cla\ is significantly smaller than \tfrag. So, even though \cla\ expands into a highly non-uniform medium with increasing density, and our age estimates are very uncertain, it is definitely too young to trigger star formation in its vicinity by a C\&C process. The dense shell around \cla\ seems to be just at the beginning of formation. Currently, \cla\ can trigger star formation only if it encounters some pre-existing dense clumps.

Triggered star formation may actually arise in three varieties in the studied region. Apart from the C\&C mechanism and compression of pre-existing clumps by the shock front around an \hii\ region, triggering by external shocks is also possible. It is important to note that the star formation in the area is not coeval. Our age estimates for the S235 and S235A \hii\ regions differ quite significantly. An age spread is seen even within the S235\,A-B cluster itself. There are three currently known massive stars in the S235\,A-B area: the exciting stars for S235\,A, S235\,B \citep{isr_felli,boley_10} and MIR-AB from \citet{dewangan_11} (`s5' source, see their Table~C8). All of them are considerably younger than lower mass stars in the same area. Ages of the lower mass stars in the cluster are about $10^5-10^6$ years \citep{dewangan_11}, while an H$_2$O maser indicates a very recent high mass star formation episode~\citep{felli_97}. This is supported by its association with the young strong HCO$^+$ outflow~\citep{felli_s235ab}. The age of the \cla\ \hii\ region is $\sim4000$~years only. Ages of MIR-AB and the exciting star in \clb\ are $\sim 3000$ and $\sim 40000$~years respectively~\citep{dewangan_11}. The age spread indicates that there were multiple star formation episodes in the vicinity of S235\,A-B.

The high-velocity gas feature FVW~172.8+1.5 probably representing an old supernova remnant~\citep{kang_koo_12} could be responsible for some of the episodes. It was found by \citet{kang_koo_12} that there is {H\,{\sc{i}}} emission with positive (from +25 to +20~\kms) and negative velocities (from --28 to --25~\kms). These negative velocities are consistent with the velocities of the molecular gas in the vicinity of S235. It means that the possible supernova is located beyond the giant molecular cloud G174+2.5 and star forming regions S235 and S235\,A-B (S235\,C also) belonging to it. We see optical emission from the \hii\ region S235 slightly obscured by interstellar gas and dust. So, S235 is situated at the near edge of the giant molecular cloud. The formation of S235 cannot be triggered by that supernova. The optical extinction toward \cla\, is about 10~mag~\citep{Krassner_79}. So, this region is more deeply embedded in the molecular gas. Streamlining of the S235\,A-B star forming region by the shock from the supernova as was modelled for example by~\citet{kimura_93} could be responsible for formation of those young stars in S235\,A-B whose age less than about 0.3~Myr~(age of the proposed SN remnant from~\citet{kang_koo_12}).

\subsection{Masses of gas and stars in the young embedded clusters}
\label{sec:dense}

Our data allows us to estimate gas masses of the dense clumps. Here we assumed that the clump borders correspond to positions where \nam\ is equal to a half of the maximum value given in Table~\ref{tab:NH3_res} for a particular clump. The gas mass is derived from $n({\rm H}_2)$ values given in Table~\ref{tab:NH3_res}. To calculate $M_{\rm star}/M_{\rm gas}$, stellar mass estimates for main sequence and pre-main sequence stars in \cltwo\ and \clcen\ from~\citet{camargo_11} are used.

We noted above that \clone\ is the densest clump in the vicinity of S235. Apparently, it is also the most massive clump in the S235 area. $M_{\rm gas}$ in \clone\ is higher than \mfrag\ for any considered value of \gnd\ (Table~\ref{tab:candcres}). No stellar mass estimate is given for \clone\ in~\citet{camargo_11} but by analogy with other clusters we may assume that it is comparable to the gas mass. In other words, the total mass of \clone\ (gas+stars) significantly exceeds the estimated mass of an average dense shell fragment formed due to the C\&C mechanism. Thus, the formation of \clone\ looks more like a result of a contraction of a pre-existing molecular clump.

Total masses (gas+stars) of \cltwo\ and \clcen\ are lower. We have seen previously that our age estimate for S235 is comparable to \tfrag\ only if the density of the surrounding gas is relatively high. In this case, clumps having appeared due to fragmentation of the shell, should have masses of the order of 50--100\,$M_\odot$. Masses of \cltwo\ and \clcen~E fall into this range, while the mass of \clcen~W is somewhat lower. It must be kept in mind that our mass estimates are quite uncertain, in particular, because the stellar component and the gas component in \cltwo\ and \clcen\ are displaced relative to each other. But still, order-of-magnitude mass estimates --- tens of solar masses for \cltwo\ and \clcen~E (vs hundreds of solar masses for \clone) --- indicate that the C\&C mechanism may have been responsible for the formation of \cltwo\ and \clcen~E. As we mentioned above, the status of \clcen~W is uncertain.

Mass is not the only parameter that distinguishes \clone\ from the other two clusters. The gas clump associated with this cluster is the only one of the considered clumps where lines of various molecular tracers are noticeably shifted relative to each other. Also, it lacks the blue CO and CS emission component that is present in \cltwo\ and \clcen\ clumps. All these differences hint at the different evolutionary track for \clone\ in comparison to \cltwo\ and \clcen\ (or, at least, to \clcen~E).

\begin{table}
\caption{Sizes and masses of dense gas in the young clumps.}
\begin{tabular}{ccccc}
\hline
Cluster     & Angular & Physical       & $M_{\rm gas}$       & $\frac{M_{\rm star}}{M_{\rm gas}}$      \\
            &  size ($^{\prime\prime}$)       & size (pc)               & ($M_\odot$)     &               \\
\hline
\clone      & 50$\times$70  & 0.4$\times$0.6 & 250 & --\\
\cltwo      & 60$\times$110 & 0.5$\times$1 & 100 & 0.4--0.6\\
\clcen\,E   & 50$\times$60 &  0.4$\times$0.5 & 40 & 0.9--1.5\\
\clcen\,W   & 50$\times$20 &  0.4$\times$0.2 & 12 & 0.9--1.5\\
\hline
\end{tabular}
\label{tab:sizeclusters}
\end{table}

\section{Conclusions}

We describe ammonia observations toward star forming regions around S235 and present maps of \am(1,1) and \am(2,2) emission. The dense gas content in young embedded clusters \clone, \cltwo, \clcen, as well as in the star forming region S235\,A-B is investigated. We determine temperature, ammonia column density and gas number density toward the clusters. Gas kinematic structure and physical parameters suggest that at least two (\cltwo\ and \clcen) of the three clusters nearest to S235 have been formed via triggering by a collect-and-collapse process if the density of the surrounding gas exceeds $3\cdot10^3$~cm$^{-3}$. The embedded cluster \clone\ could have been formed as a result of an interaction of the shock front from S235 with a pre-existing dense clump. The new observations are consistent with our previous conclusion that \clone\ is at an earlier evolutionary stage than \cltwo\ and \clcen. Parameters of the dense gas toward \cla\ imply that this \hii\ region expands into a medium with increasing density. The \cla\ \hii\ region is too young to trigger star formation around itself by a collect-and-collapse process. However, it can trigger formation of stars in pre-existing clumps. External large-scale shocks could have been responsible for shaping the S235\,A-B star forming region.

CO, CS, and NH$_3$ lines are shifted relative to each other in \clone. A shift between CO and NH$_3$ lines confirm our conclusion from Paper~I that there is a relative motion between dense and diffuse gas in this object. We note also that star formation in S235\,A-B is not coeval. Massive stars in the region are considerably younger than the lower mass stars in the region.

\section*{Acknowledgments}

This work was supported by the RFBR (grants 13-02-00642, 11-02-01332) and by the President of the RF grant NSh-3602.2012.2. MSK and DSW are grateful to the OFN-17 program of the Russian Academy of Sciences. MSK and AMS thank the staff of the Effelsberg Observatory for their hospitality and assistance during observations. We thank Aurora Sicilia-Aguilar for help in observations of \cla.

\label{lastpage}

\end{document}